\documentclass[aps,reprint,showpacs,superscriptaddress,floatfix]{revtex4-1}
\pdfoutput=1
\usepackage{amsmath,amssymb,amsfonts}
\usepackage{hyperref}
\usepackage{multirow}
\usepackage{graphicx}
\usepackage[caption=false]{subfig}
\usepackage{caption,setspace}
\usepackage{dsfont}
\usepackage{algorithmic}
\usepackage{textcomp}

\DeclareMathOperator{\Tr}{Tr}
\newcommand{\MPQ}{Max-Planck-Institut f\"ur Quantenoptik, Hans-Kopfermann-Str.\ 1, D-85748 Garching, Germany}
\newcommand{\MCQST}{Munich Center for Quantum Science and Technology (MCQST), Schellingstr.\ 4, 80799 M\"unchen, Germany}

\begin{document}

\title{From probabilistic graphical models to generalized tensor networks\\ for supervised learning}

\author{Ivan Glasser}
\affiliation{\MPQ}
\affiliation{\MCQST}
\author{Nicola Pancotti}
\affiliation{\MPQ}
\affiliation{\MCQST}
\author{J. Ignacio Cirac}
\affiliation{\MPQ}
\affiliation{\MCQST}

\begin{abstract}
Tensor networks have found a wide use in a variety of applications in physics and computer science, recently leading to both theoretical insights as well as practical algorithms in machine learning. In this work we explore the connection between tensor networks and probabilistic graphical models, and show that it motivates the definition of generalized tensor networks where information from a tensor can be copied and reused in other parts of the network. We discuss the relationship between generalized tensor network architectures used in quantum physics, such as string-bond states, and architectures commonly used in machine learning. We provide an algorithm to train these networks in a supervised-learning context and show that they overcome the limitations of regular tensor networks in higher dimensions, while keeping the computation efficient. A method to combine neural networks and tensor networks as part of a common deep learning architecture is also introduced. We benchmark our algorithm for several generalized tensor network architectures on the task of classifying images and sounds, and show that they outperform previously introduced tensor-network algorithms. The models we consider also have a natural implementation on a quantum computer and may guide the development of near-term quantum machine learning architectures.
\end{abstract}

\maketitle

\section{Introduction}
\label{sec:introduction}

Tensor networks, which factorize a high-order tensor into a network of low-order tensors, have found a wide use of applications from quantum physics \cite{Orus2014,VerstraeteReview} to machine learning \cite{MAL-059,MAL-067}. They can be used to compress weights of neural networks \cite{Novikov2015,627946391,627946323,627946324}, to study model expressivity \cite{627946320,Cohen2016,627946390,Cohen2017,Levine2017,Glasser2019} or to parametrize complex dependencies between variables \cite{627946317,Stoudenmire2016,Novikov2016,627946325}. Recently, they have also attracted attention in the context of quantum machine learning \cite{BiamonteReview,Dunjko2018}: there has been much interest in understanding how low-depth quantum circuits that can be implemented on near-term quantum devices may be useful in machine learning \cite{Farhi2014,McClean2016,Farhi2018,Schuld2018,Benedetti2018,Mitarai2018}, and tensor networks are a natural tool to perform the classical simulation of such algorithms \cite{Huggins2018,Grant2018,Glasser2019}. Tensor networks that can be efficiently contracted on classical computers thus provide a unique platform to benchmark and guide the development of new quantum machine learning architectures.

In this work, we explore the relationship between tensor networks and more common machine learning architectures, in particular probabilistic graphical models. We define generalized tensor networks which connect the two frameworks. These networks rely on the copy and reuse of local tensor information. Unlike regular tensor networks, they can be defined in complex geometries while remaining efficient to contract as long as an appropriate hierarchical order can be defined. We apply several variants of generalized tensor networks to image classification and environmental sound recognition and compare their performance, concluding that generalized tensor networks typically perform better than tensor networks alone.
 
Since they use several copies of the data inputs, generalized tensor networks share some structure with convolutional neural networks (CNN) \cite{Fukushima1979,LeCun1989}, while having direct connections to restricted Boltzmann machines (RBM) \cite{Smolensky1986,Hinton2002}. Examples of such networks that have been used in quantum physics include string-bond states (SBS) \cite{SBS2008,Sfondrini2010} (that generalize matrix product states (MPS)/tensor-trains \cite{Oseledets2011})  as well as entangled plaquette states (EPS) \cite{Gendiar2002,Mezzacapo2009,Changlani2009}. Particular generalized tensor networks with tree structures have also been used to study the expressivity of deep learning models \cite{Cohen2016,Levine2018}.

We introduce an algorithm for performing supervised learning with generalized tensor networks, which combines stochastic gradient descent with previously introduced approaches for tensor networks. This framework generalizes works based on regular tensor networks such as MPS \cite{Stoudenmire2016,Novikov2016,Han2017,Liu2018} or tree tensor networks \cite{Liu2017}. It has the advantage that more complex structures can be formed while keeping the computation efficient. This is especially useful for data that possesses some geometrical structure in more than one dimension, such as images. We emphasize that the algorithm does not need to rely on any Monte Carlo techniques, unlike in quantum physics, where generalized tensor networks can only be optimized in combination with computationally expensive Monte Carlo sampling. In particular the cost of optimizing a SBS is only a constant factor times the cost of optimizing a MPS, but SBS are much more flexible in higher dimensions and can interpolate between a MPS and a restricted Boltzmann machine.

We discuss how real-valued data can be used in conjunction with tensor networks and suggest to learn the relevant tensor features of real data as part of the network. Inspired by deep network architectures, we also propose two ideas to combine neural networks and tensor networks. In the first case we use a neural network to extract features from the data in order to feed them into a tensor network, in the second we combine generalized tensor networks and neural networks in the same deep network architecture.

We benchmark our algorithms for several generalized tensor network architectures on different data sets \cite{fashionmnistreference,MNISTreference,Piczak2015}. For image classification, we find that generalized tensor networks outperform previously introduced tensor-network algorithms based on MPS or trees while keeping a small dimension of the tensors. In the context of environmental sound recognition, we find that MPS and SBS deliver comparable performance. This shows that SBS should also be used along with MPS when considering one-dimensional data, especially in the presence of long-range correlations. Furthermore they may be applied in other settings such as natural language processing.

The network architectures we consider have a natural implementation on a quantum computer, which shows that simple quantum circuits that can be simulated classically can already achieve a good performance in supervised learning. It is still an open question whether quantum circuits that cannot be simulated classically will provide an advantage over classical machine-learning techniques. The networks we introduce may thus serve as a natural tool to test the performance and guide the development of such circuits.

\section{Graphical Models and Generalized Tensor Networks}
\label{section1}

We first review definitions of probabilistic graphical models and tensor networks, discuss their relationship and show that the two frameworks can be connected through the definition of generalized tensor networks in which parts of the network can be copied and reused. Examples of generalized tensor networks which have been successfully used in quantum physics are introduced, and their connection to more common machine-learning architectures is discussed.

\subsection{Graphical models}

\begin{figure}[t]
\centering
\subfloat[]{\includegraphics[width = 4cm]{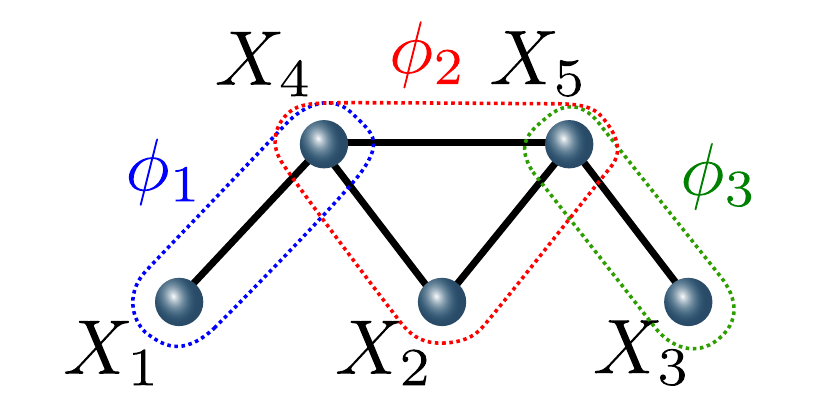}
\label{undirectedgraphmodel}} 
\subfloat[]{\includegraphics[width = 4cm]{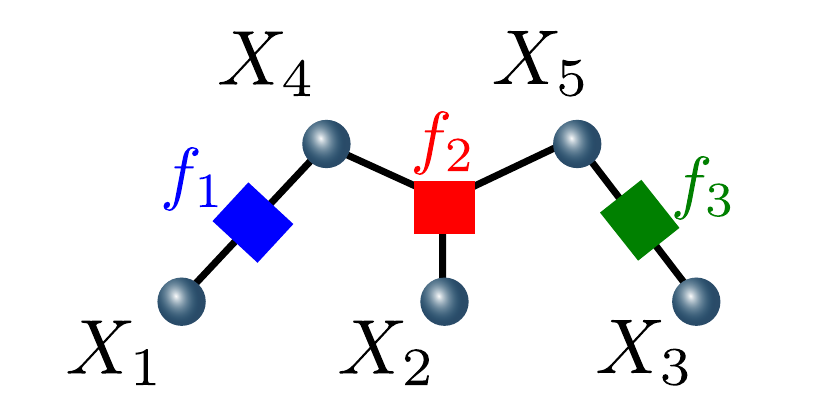} \label{factorgraph}} \\
\subfloat[]{\includegraphics[width = 4cm]{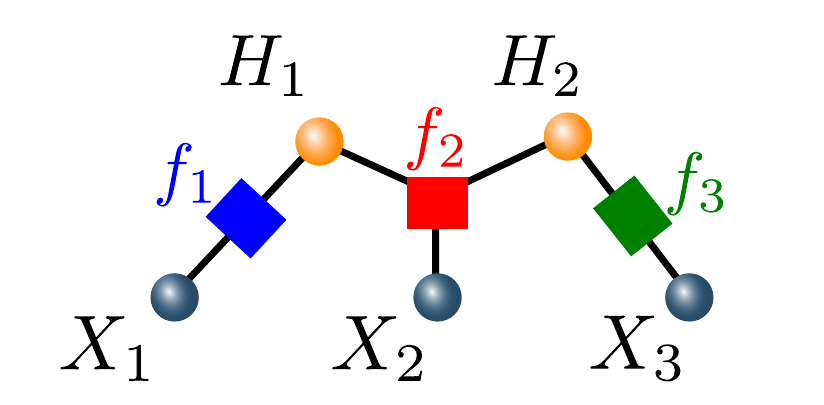} \label{factorgraphmarginalized}}
\subfloat[]{\includegraphics[width = 4 cm]{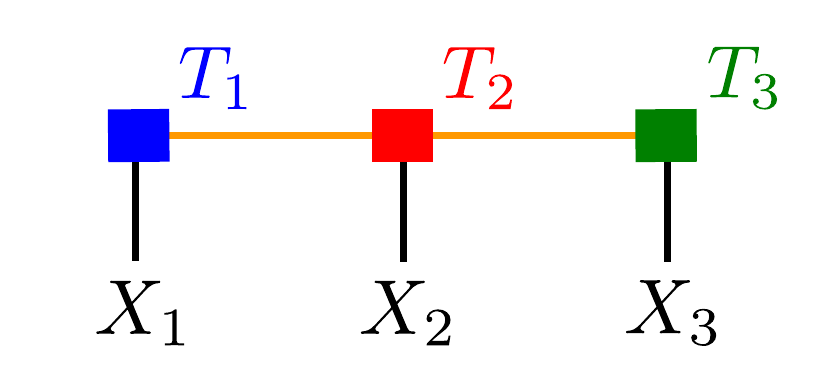} \label{mps}}
\caption{(a) Undirected graphical model with three maximal cliques depicted in colors (b) Corresponding factor graph (c) Factor graph with hidden units in orange that are marginalized (d) Equivalent tensor network, which is a Matrix Product State.}
\end{figure}

Let us consider a set of discrete random variables $\mathbf{X}=\{X_1,\ldots,X_N\}$ taking values $\mathbf{x}=(x_1,\ldots,x_N)$ and a data set of samples from these variables $\mathcal{D}=\{\mathbf{d}_1,\ldots,\mathbf{d}_{|D|}\}$. Inferring the underlying probability distribution $p(\mathbf{x})$ can be done by maximizing the log-likelihood 
\begin{align}
\mathcal{L}=\sum_{i=1}^{|D|} \log p(\mathbf{d}_i).
\end{align}
A common choice of parametrized models for $p$ are graphical models \cite{graphicalmodels}, which correspond to a factorization of the probability distribution over a graph. Consider a graph $G=(V,E)$, where $V$ is a set of vertices, $E$ a set of edges between these vertices (each $e\in E$ is a pair of elements in V) and $\text{cl}(G)$ is the set of maximal cliques of the graph. An undirected graphical model or a Markov random field defines a factorization of the joint probability of all random variables as
\begin{align}
p(\mathbf{X}=\mathbf{x})=\frac{1}{Z}\prod_{C\in \text{cl}(G)}\phi_C(\mathbf{x}_C),
\end{align}
where $\mathbf{x}_C$ are the values of the variables in clique $C$, $\phi_C$ are the clique potentials which are positive functions and $Z$ is the partition function that ensures normalization of the probability (Fig.~\ref{undirectedgraphmodel}). Graphical models can be converted to factor graphs \cite{Frey2003} defined on a bipartite graph of factors and variable vertices: one factor node $f_C$ is created for each maximal clique and the factor is connected to the variables in the corresponding clique (Fig.~\ref{factorgraph}). The factorization of the probability distribution still reads
\begin{align}
p(\mathbf{X}=\mathbf{x})=\frac{1}{Z}\prod_{C}f_C(\mathbf{x}_c),
\end{align}
and inference can be performed through belief propagation and the sum-product algorithm on factor graphs. To increase the set of distributions which can be represented we can add additional dependencies by introducing ancillary hidden variables (which are unobserved, that is their values are not supplied in the data) $\mathbf{H}=\{H_1,\ldots,H_M\}$ (Fig.~\ref{factorgraphmarginalized}). The resulting probability distribution is obtained by marginalizing these hidden variables, giving
\begin{align}
p(\mathbf{X}=\mathbf{x})=\frac{1}{Z}\sum_{\mathbf{h}}\prod_{C}f_C(\mathbf{x}_C,\mathbf{h}_C).
\end{align}

\subsection{Tensor networks}

\begin{figure}[t]
 \centering
 \includegraphics[width = 8cm]{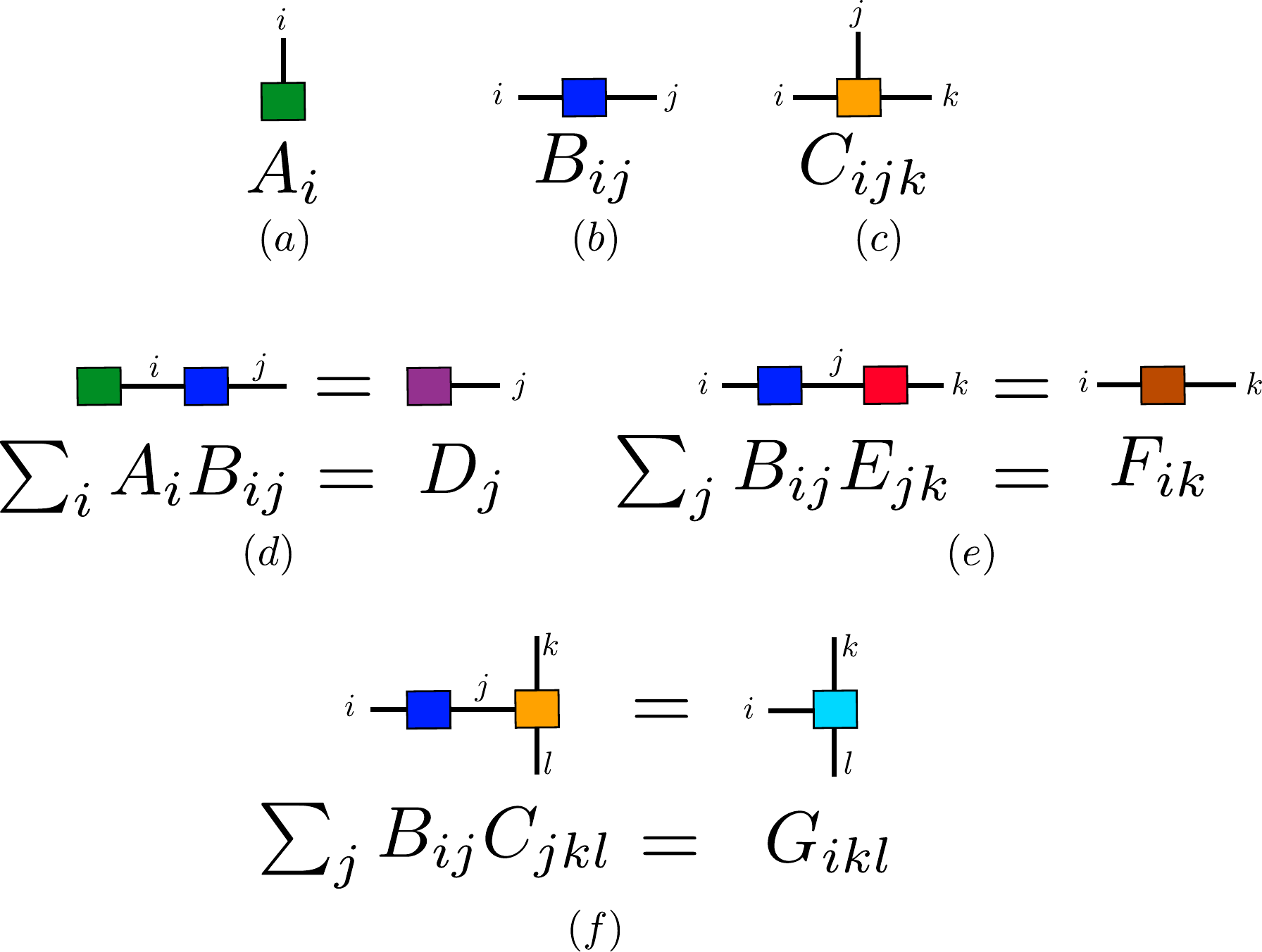}
 \caption{Graphical notation for tensor networks : (a) vector, (b) matrix, (c) order 3 tensor, (d) vector-matrix multiplication (e) matrix-matrix multiplication (f) matrix-tensor contraction.}
 \label{tensorcontractions}
 \end{figure}

Tensor networks are factorizations of tensors (multi-dimensional arrays) into a network of smaller tensors.  This network admits a graphical notation, depicted in Fig.~\ref{tensorcontractions}, in which boxes represent tensors and legs coming out of these boxes represent tensor indices. A leg joining two tensors represents a sum (contraction) over a joint index, such that the resulting tensor is a sum over joint indices of products of tensors. A simple example is a matrix factorization, in which a matrix is factorized in a product of two matrices. More formally, we consider a graph $G=(V,E)$ where we add to $E$ edges containing only one vertex. These will correspond to open legs in a tensor-network graphical representation. We denote $E'$ the subset of $E$ containing edges that connect two vertices. We associate an integer size $D_e$ called the bond dimension to each edge and define a tensor $T_v\in\otimes_{e\in v} \mathbb{R}^{D_e}$ for each vertex $v \in V$, with indices associated with the edges of this vertex. A tensor-network state is defined by summing over indices on all edges in the graph that connect two tensors. The resulting contracted tensor network is a tensor indexed by the indices of the open legs, denoted as $\mathbf{x}=(x_1,\ldots,x_N)$:
\begin{align}\label{eq:tensornetwork}
T_{\mathbf{x}}=\sum_{e \in E'} \prod_{v}T_v.
\end{align}

A particular case are one-dimensional matrix product states (MPS) (Fig.~\ref{mps}), also known as tensor-trains \cite{Oseledets2011}, which decompose a tensor as
\begin{align}
T_{x_1,\ldots,x_N}=\sum_{e\in E'} A^{x_1}_{e_1}A^{x_2}_{e_2,e_3}A^{x_3}_{e_3,e_4}\cdots A^{x_N}_{e_N},
\end{align}
where, for fixed value of $\mathbf{x}$, $A^{x_1}$ and $A^{x_N}$ are vectors, and $A^{x_j}$, $j=2,\ldots,N-1$ are matrices. On a closed chain, also known as tensor ring \cite{Zhao2017}, the corresponding decomposition is
\begin{align}
T_{x_1,\ldots,x_N}=\Tr\left(\prod_{j=1,\ldots,N} A^{x_j}\right),
\end{align}
where all $A^{x_j}$ are matrices, and the dimension of these matrices is the bond dimension. Generalizations to trees and lattices in higher dimensions have also been studied.

\subsection{Tensor networks and graphical models duality}

Consider a graphical model given as a factor graph in the special case where each hidden variable is connected to two factors, and each visible variable connected to one factor, as is the case in Fig.~\ref{factorgraphmarginalized}. Because all the variables are discrete, each factor $f_C$ is a non-negative function of discrete variables to which it is connected. It can therefore be written as a tensor $F_{C,\mathbf{x}_C,\mathbf{h}_C}$ with non-negative entries indexed by the variables $\mathbf{x}_C$ and $\mathbf{h}_C$ to which it is connected. The factor graph then defines a probability mass function over the visible variables that can be represented by a non-negative tensor $P_\mathbf{x}$ satisfying
\begin{align}
P_\mathbf{x}=\sum_{\mathbf{h}}\prod_{C}F_{C,\mathbf{x}_C,\mathbf{h}_C}.
\end{align}
By comparing this expression with \eqref{eq:tensornetwork}, we see that this probability mass function can be interpreted as a tensor-network state. Marginalization of a hidden variable corresponds to contracting the indices of the different factors connected to this variable, as indicated by the orange lines in Fig.~\ref{mps}, and the visible variables correspond to the open legs of the tensor network. In the following we will refer to these open legs that correspond to visible variables or data inputs as the inputs of the tensor network. Probabilistic graphical models with discrete variables are therefore tensor networks of non-negative tensors. This connection has been previously observed in particular models \cite{Morton2014,Kliesch2014,Chen2017,Glasser2017}, and \cite{Robeva2017} provides a more detailed analysis of this duality on hypergraphs. 

The fact that graphical models rely on non-negative tensors, whereas tensor networks are usually studied in the context of real (or complex) elements, has important consequences for the optimization algorithms. Graphical models can be used in conjunction with expectation-maximization algorithms, which rely on the computation of conditional probabilities over a subset of the variables. In general, tensor networks cannot use the same algorithms, since they do not have the same probabilistic interpretation. However, they can use other powerful optimization algorithms, such as the density matrix renormalization group (DMRG). These algorithms can rely on the singular-value decomposition of real and complex matrices, without the constraint of non-negativity. A comparison of the expressive power of non-negative tensor networks and real or complex tensor networks in the context of probabilistic modeling has been performed in \cite{Glasser2019}.

\subsection{Duality with copy tensors}

\begin{figure}[t]
\centering
\subfloat[]{\includegraphics[width = 4cm]{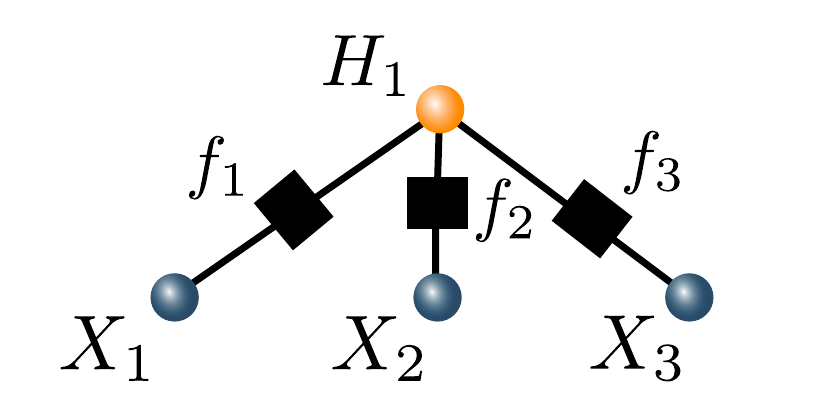}
\label{factorgraphcopyhidden}} 
\subfloat[]{\includegraphics[width = 4cm]{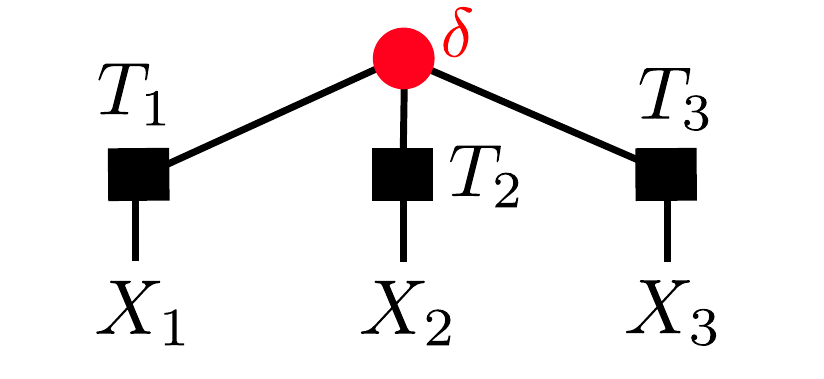} \label{factorgraphcopyhiddentensor}} \\
\subfloat[]{\includegraphics[width = 4cm]{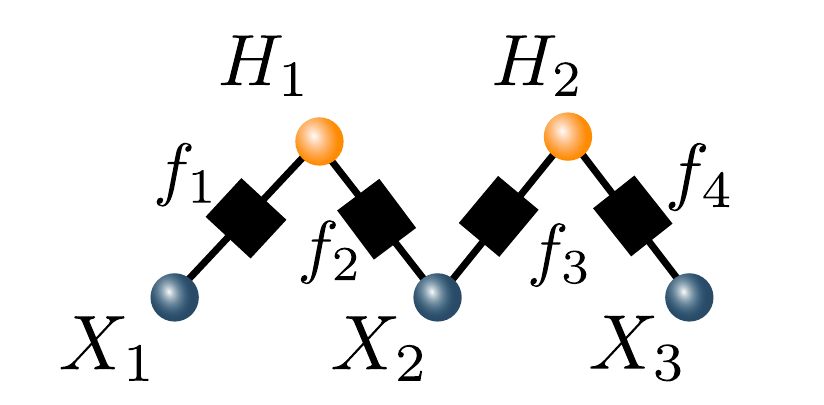} \label{factorgraphcopyvisible}}
\subfloat[]{\includegraphics[width = 4 cm]{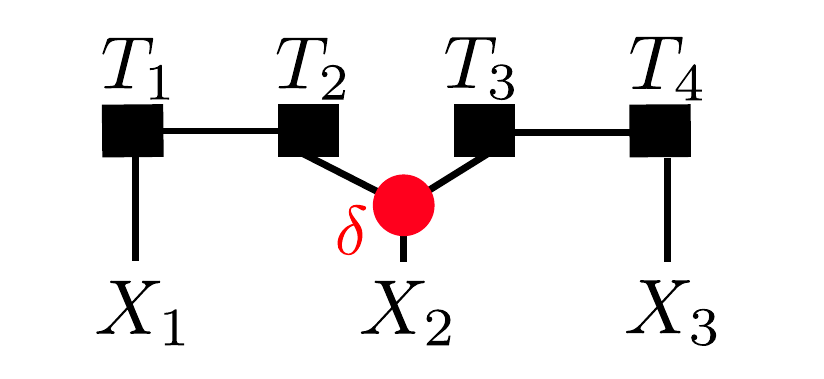} \label{factorgraphcopyvisibletensor}}\\
\subfloat[]{\includegraphics[width = 4 cm]{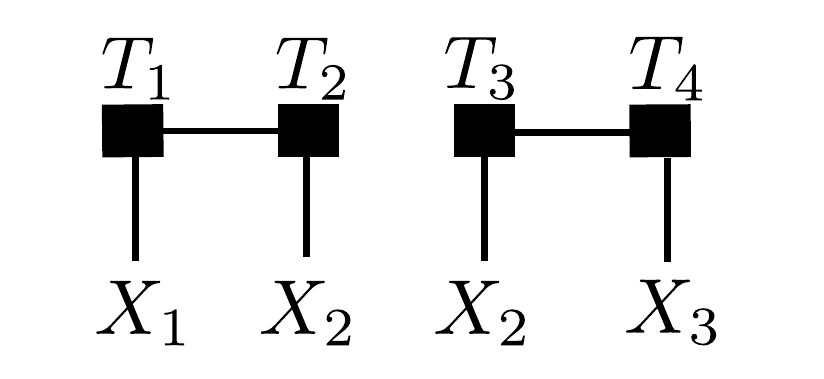} \label{factorgraphcopyvisibletensor2}}
\caption{(a) Factor graph with one hidden variable connected to several factors (b) Equivalent tensor network with copy tensor (c) Factor graph with visible variables connected to several factors (d) Equivalent tensor network (e) Corresponding tensor network if the visible variables have fixed input values.}
\end{figure}

Consider now the case of a probabilistic graphical model where one hidden variable is connected to several factors, as in Fig.~\ref{factorgraphcopyhidden}. To write the resulting probability mass function as a tensor network, we need to sum over the values of the hidden variable, which corresponds to summing over the value of the corresponding index which appears in several tensors. To represent this operation as a simple tensor network, we introduce a copy tensor $\delta$ with a number of indices corresponding to the number of factors connected to the hidden variable (each index can take as many values as the hidden variable), and such that $\delta$ is equal to one if all indices take the same value, and zero otherwise \cite{Biamonte2011,Clark2017}. In the case of Fig.~\ref{factorgraphcopyhidden}, $\delta$ is an order-3 tensor such that $\delta_{ijk}=1$ if $i=j=k$ and $\delta_{ijk}=0$ otherwise. We represent this tensor as a red dot in graphical representation. By inserting this copy tensor in the tensor network and connecting it to the corresponding factors, we obtain that the probability mass function defined by the graphical model can be represented by a tensor network, as in Fig.~\ref{factorgraphcopyhiddentensor}.

The same procedure can be used if a visible variable is connected to several factors, as in Fig.~\ref{factorgraphcopyvisible} and Fig.~\ref{factorgraphcopyvisibletensor}. Note that when computing the probability of inputs $\mathbf{x}$, the visible variables take specific given values. In this case contracting an input with a copy tensor will give the same result as fixing the value of all legs connected to the copy tensor to the input value and contracting the rest of the tensor network, as in Fig.~\ref{factorgraphcopyvisibletensor2}.

\subsection{Application to restricted Boltzmann machines}

\begin{figure}[t]
\centering
\subfloat[]{\includegraphics[width = 4cm]{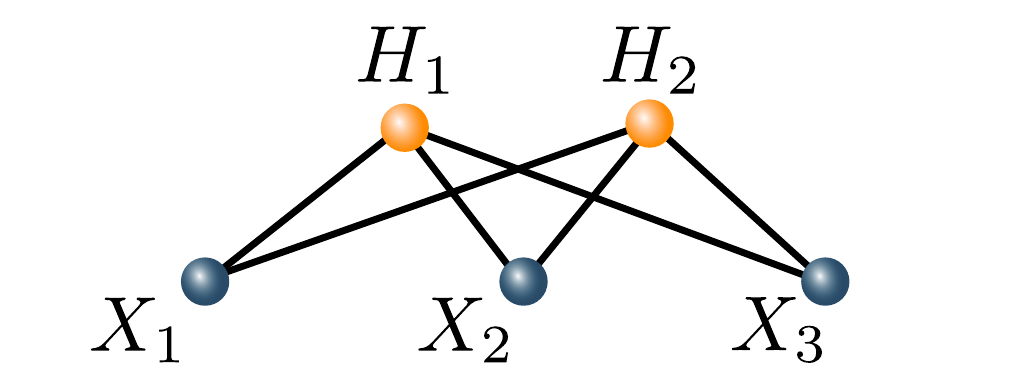}\label{RBM3}}
\subfloat[]{\includegraphics[width = 4cm]{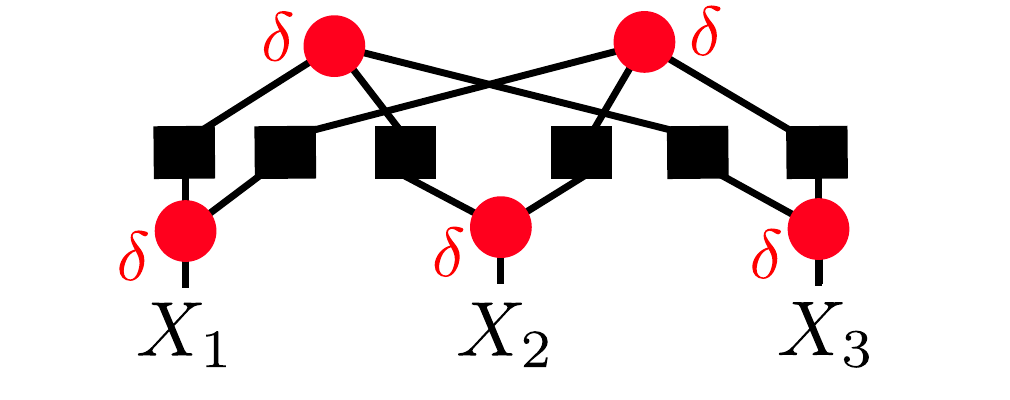}\label{RBM3tensor}}\\
\subfloat[]{\includegraphics[width = 4cm]{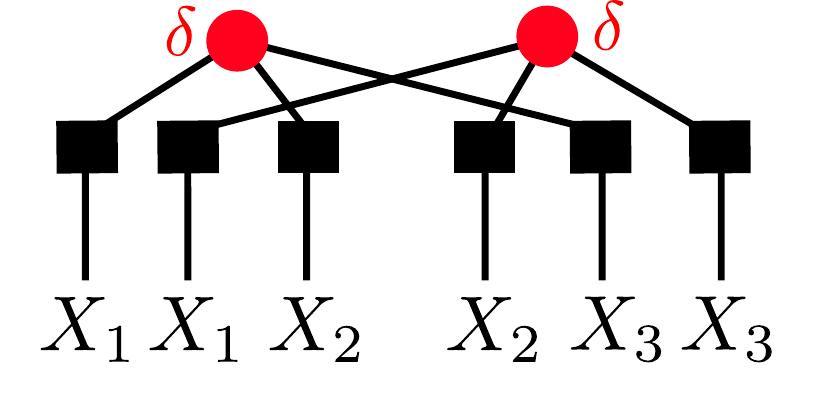}\label{RBM3input}}
\subfloat[]{\includegraphics[width = 4cm]{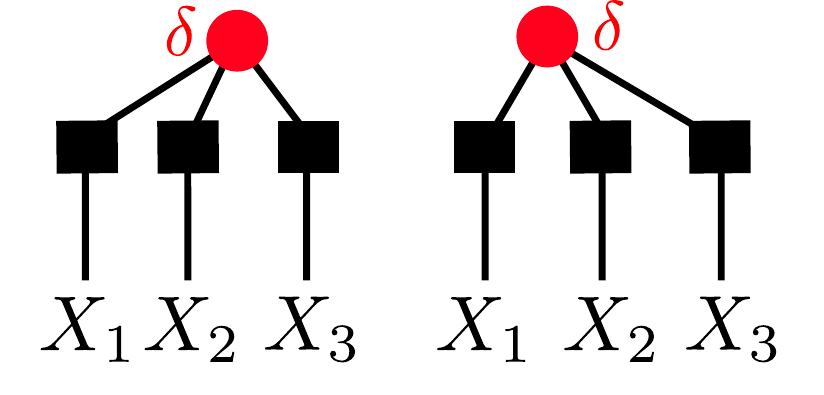}\label{RBM3input2}}
\caption{(a) Restricted Boltzmann machine (RBM) (b) Corresponding tensor network (c) Corresponding tensor network when the input values are fixed and the input copy tensors contracted  (d) Same tensor network with a reordering of the inputs and tensors.}
\end{figure}

One class of probabilistic graphical model which has the property that visible and hidden variables are connected to several factors are restricted Boltzmann machines (RBM) \cite{Smolensky1986,Hinton2002}. They are defined on a bipartite graph with visible variables $\mathbf{X}$ and hidden variables $\mathbf{H}$ (Fig.~\ref{RBM3}), so each visible variable is connected to each hidden variable. The connections between variables on this graph take the form of Ising interactions and the probability distribution of joint variables is
\begin{align}
p(\mathbf{x},\mathbf{h})=\frac{1}{Z} e^{\mathcal{H}(\mathbf{x},\mathbf{h})},
\end{align}
where the Hamiltonian $\mathcal{H}$ is a classical Ising Hamiltonian defined as (we omit here the bias terms for simplicity) 
\begin{align}
\mathcal{H}=\sum_{i,j} w_{ij} h_i x_j.
\end{align}
In the case where both visible and hidden variables are binary valued, the resulting probability distribution once the hidden variables have been marginalized is 
\begin{align}
p(\mathbf{x})&=\frac{1}{Z} \sum_{\mathbf{h}}e^{\mathcal{H}(\mathbf{x},\mathbf{h})}, \\
&=\frac{1}{Z} \prod_i (1+e^{\sum_j w_{ij} x_j }).\label{rbmfunction}
\end{align}
Let us represent the corresponding factor graph, and associated tensor network. The cliques of the graph are all pairs of visible and hidden variables, for which there is an associated factor, which is a non-negative tensor, $f(x_j,h_i)=e^{w_{ij} h_i x_j}$. By inserting a copy tensor at the position of the hidden variables, as well as of the input variables, we obtain the tensor network representation of the RBM in Fig.~\ref{RBM3tensor} \cite{Clark2017}.

\subsection{Relationship with models used in quantum physics}

If we look at the graph for the tensor network equivalent to the restricted Boltzmann machine in Fig.~\ref{RBM3tensor}, we observe that this graph has many loops. Nevertheless, one can efficiently and exactly compute $p(\mathbf{x})$ for arbitrary size using \eqref{rbmfunction}. This comes from the fact that if we fix the values of the input variables and contract the copy tensors connected to the inputs, then the corresponding tensor network has no loops, as depicted in Fig.~\ref{RBM3input2}.

\begin{figure}[t]
\centering
\subfloat[RBM]{\includegraphics[width = 4cm]{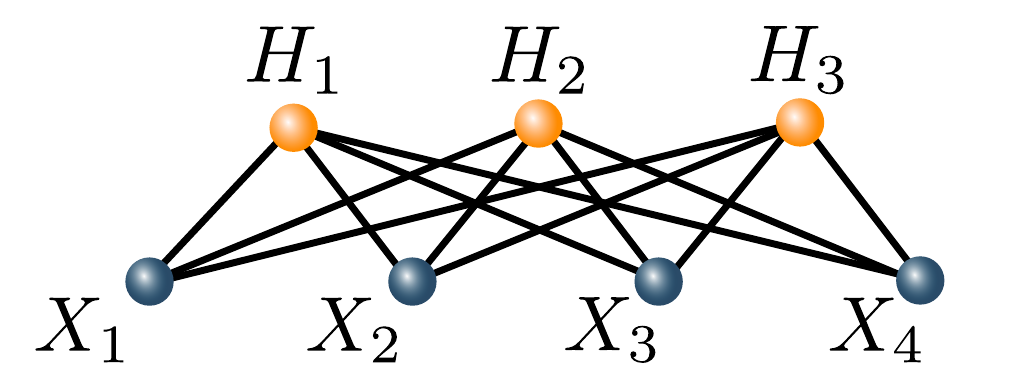}
\label{RBM}} 
\subfloat[SBS]{\includegraphics[width = 4cm]{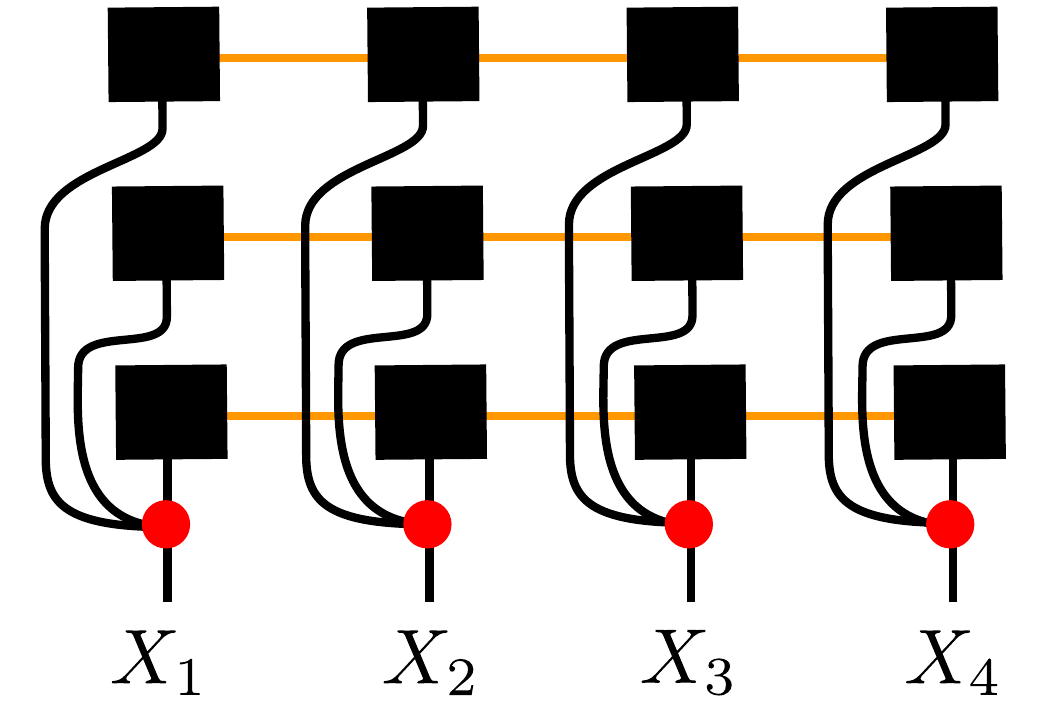} \label{SBS}} \\
\subfloat[Short-range RBM]{\includegraphics[width = 4cm]{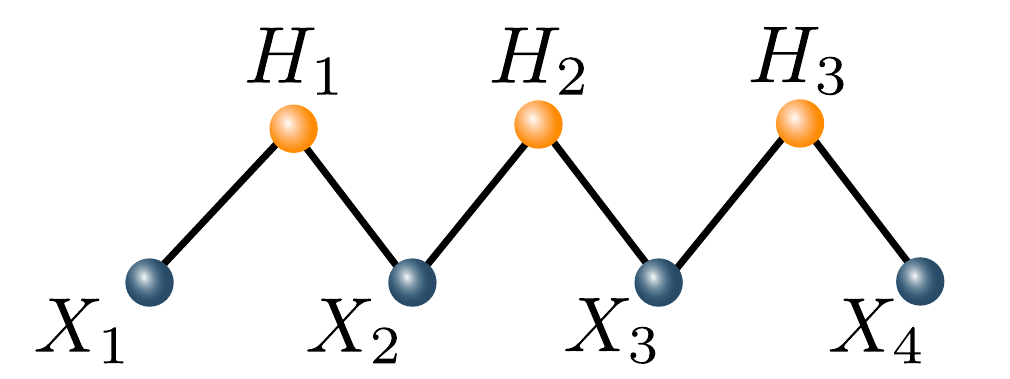} \label{short-rangeRBM}}
\subfloat[EPS]{\includegraphics[width = 4 cm]{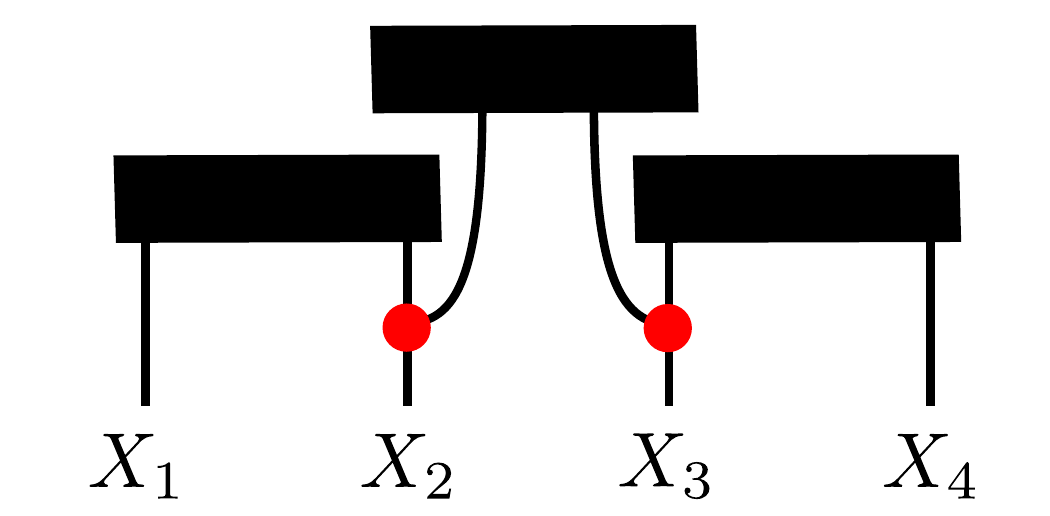} \label{EPS}}
\caption{(a) Restricted Boltzmann machine (RBM) consisting of visible and hidden variables (b) String-bond state with 1D geometry generalizing RBM. The legs corresponding to contracted indices in each MPS are depicted in orange for visibility. (c) Short-range RBM with local connections between visible and hidden variables (d) Entangled plaquette state (EPS) generalizing the short-range RBM.}
\end{figure}

Particular architectures of tensor networks satisfying the same property are tree tensor networks with copied inputs \cite{Cohen2016,Levine2018}. Other examples of such networks have been used in the quantum physics community. The simplest example are entangled plaquette states (EPS) \cite{Gendiar2002,Mezzacapo2009,Changlani2009}, also known as correlator product states, in which the tensor network is defined as a product of smaller tensors on overlapping clusters of variables:
\begin{align}
T_{x_1,\ldots,x_N}=\prod_{p=1}^{P} T_p^{\mathbf{x}_p},
\end{align} 
where a coefficient $T_p^{\mathbf{x}_p}$ is assigned to each of the $2^{n_p}$ (for binary variables) configurations $\mathbf{x}_p$ of the variables in cluster $p$. Because the clusters overlap, the value of each variable is copied to all the tensors in which it is included (Fig.~\ref{EPS}). A sparse or short-range RBM is a special case of EPS \cite{Glasser2017,Chen2017}, in which the tensor $T_p^{\mathbf{x}_p}$ takes the particular form $(1+e^{\sum_j w_{pj} x_j })$, where the sum is limited to the variables in each cluster.

Another example are string-bond states (SBS) \cite{SBS2008,Sfondrini2010}, defined by placing MPS over strings (each string $s$ is an ordered subset of the set of variables) on a graph which does not need to be a one-dimensional lattice. The resulting tensor network is 
\begin{align}
T_{x_1,\ldots,x_N}=\prod_{s} \Tr\left(\prod_{j\in s} A^{x_j}_{s,j} \right).
\end{align} 
The value of each visible variable is copied and sent to different MPS (Fig.~\ref{SBS}). It was shown in \cite{Glasser2017} that a RBM is a special case of SBS for which each string is associated with a hidden variable and covers all visible variables, and the matrices are taken to be
\begin{align}
A^{x_j}_{s,j}=\begin{pmatrix}
1 & 0 \\
0 & e^{w_{sj} x_j}
\end{pmatrix}.
\end{align}
SBS thus provide a generalization of RBM that is naturally defined for discrete variables of any dimension and can introduce different correlations through the use of higher dimensional and non-commuting matrices. Since SBS also include a MPS as a particular case, they provide a way to interpolate between a MPS (large bond dimension, only one string) and a RBM (bond dimension 2, diagonal matrices, many strings). Note that a product of $S$ MPS of bond dimension $D$ is an MPS with bond dimension $D^S$, so restricted Boltzmann machines can be written as MPS of large bond dimension but with diagonal matrices.

\subsection{Generalized Tensor Networks}

\subsubsection{Input features for tensor networks}

\begin{figure}[t]
\centering
\includegraphics[width = 4cm]{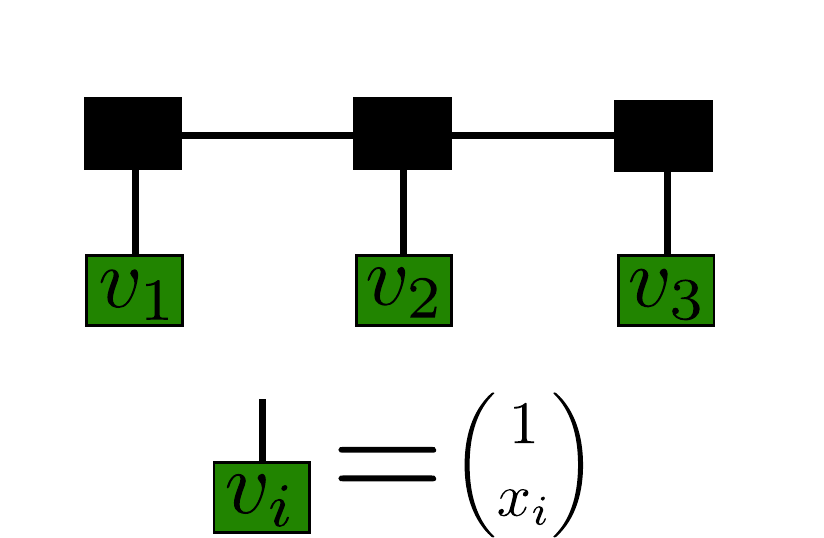}
\caption{MPS with feature vectors as inputs.}
\label{fig:featuresinput}
\end{figure}

In many cases of interest, data is not discrete, but instead given in the form of real numbers. In order to train tensor networks to perform machine learning tasks on such data, it has been suggested to define feature vectors from the data, and then to contract these feature vectors with a tensor network to define a function of the inputs \cite{Stoudenmire2016,Novikov2016}. Consider for example input data given as real numbers $\{x_1,\ldots,x_N\}$ and define feature vectors 
\begin{align}
v_i= \begin{pmatrix}
1 \\ x_i
\end{pmatrix}.
\end{align}
Now it is possible to use a tensor network to define a function of the inputs by contracting these vectors with the open legs of a tensor network, as depicted in Fig.~\ref{fig:featuresinput} in the case of a MPS.

\subsubsection{Copy operation with vector inputs}

We would like to extend the previous results that use copy tensors to copy the inputs of the tensor network, allowing to efficiently contract complex tensor networks when the inputs are given. This idea that input data may be copied is also used in other machine learning architectures, such as for example Convolutional Neural Networks (CNN): a convolutional layer applies several filters to local clusters of input variables. In order to apply different filters to the same input data, this data needs to be copied and fed to the different parts of the neural network. In the case of tensor networks, we face the problem that copy tensors can only be used to copy discrete inputs \cite{Levine2018}. Placing copy tensors as the input of the tensor network will therefore not have the expected effect.

For these reasons we introduce a different notion of copy, that we call the copy operation, that allows to copy real vectors. We will call the models that use this feature generalized tensor networks, to distinguish them from tensor networks with copy tensors, and we will restrict ourselves to the special cases in which these networks can be efficiently contracted as long as the inputs have fixed values. In the special case where the copy only takes place at the level of the input variables, and that these variables are discrete, then these networks will coincide with tensor networks that use a copy tensor at the level of the input legs.

\begin{figure}[t]
\centering
\subfloat[]{\includegraphics[width = 4cm]{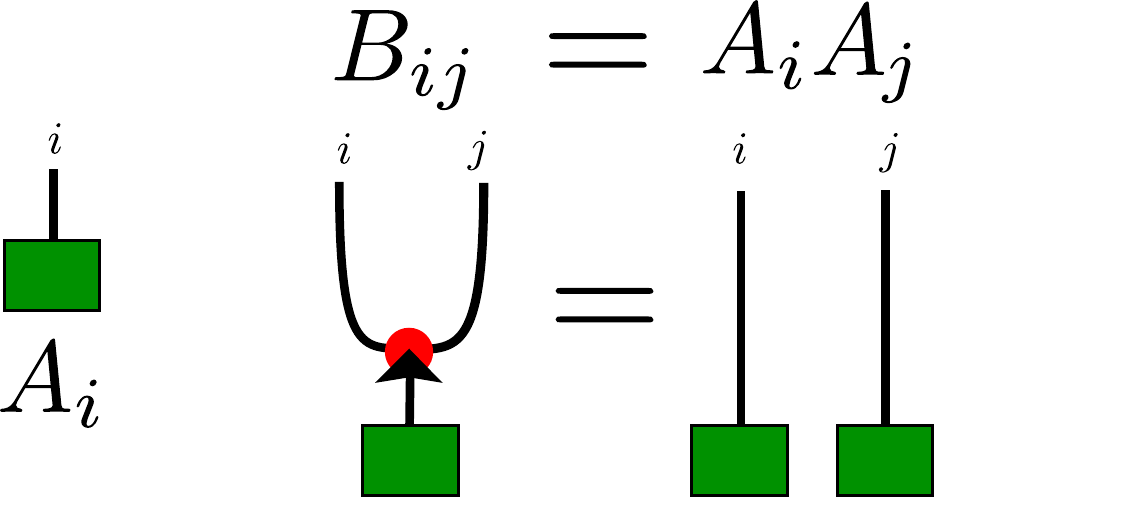}}
\subfloat[]{\includegraphics[width = 4cm]{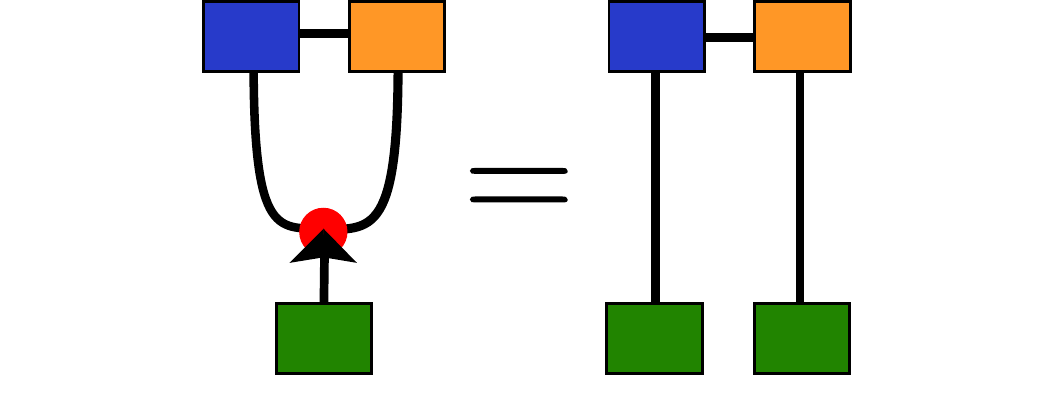}}\\
\subfloat[]{\includegraphics[width = 8cm]{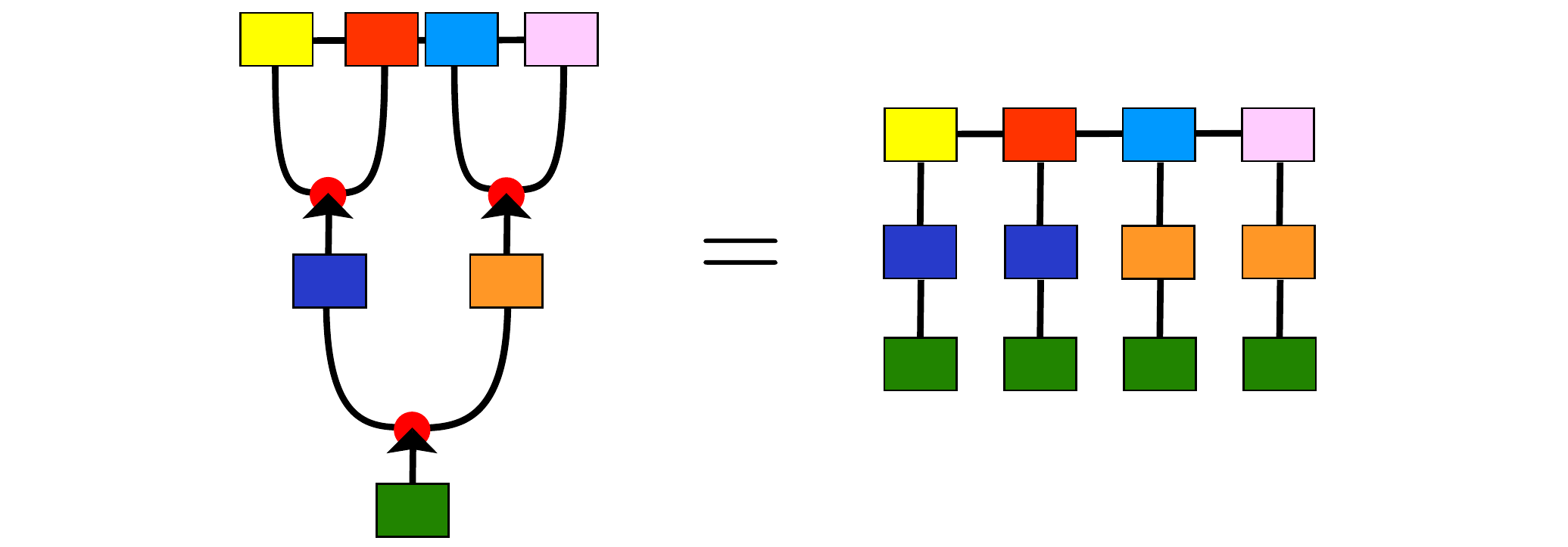}}\\
\subfloat[]{\includegraphics[width = 8cm]{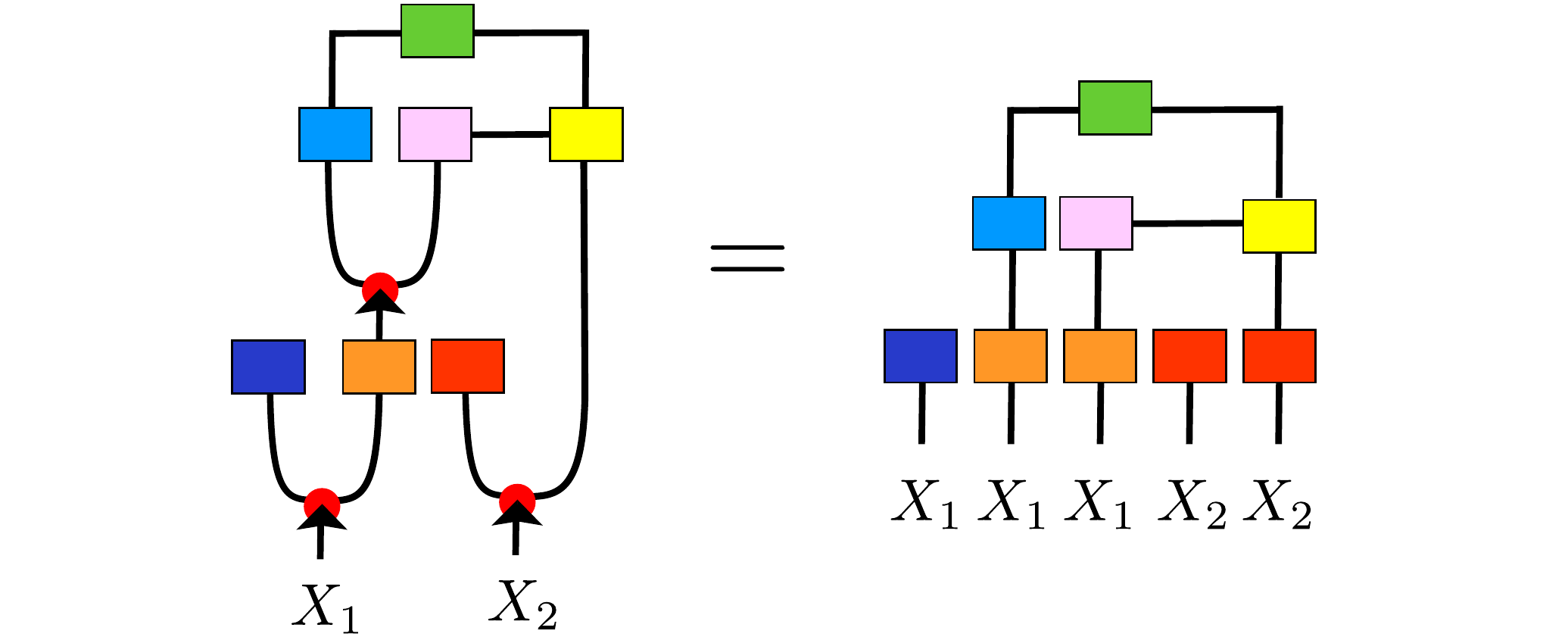}}\\
\subfloat[]{\includegraphics[width = 8cm]{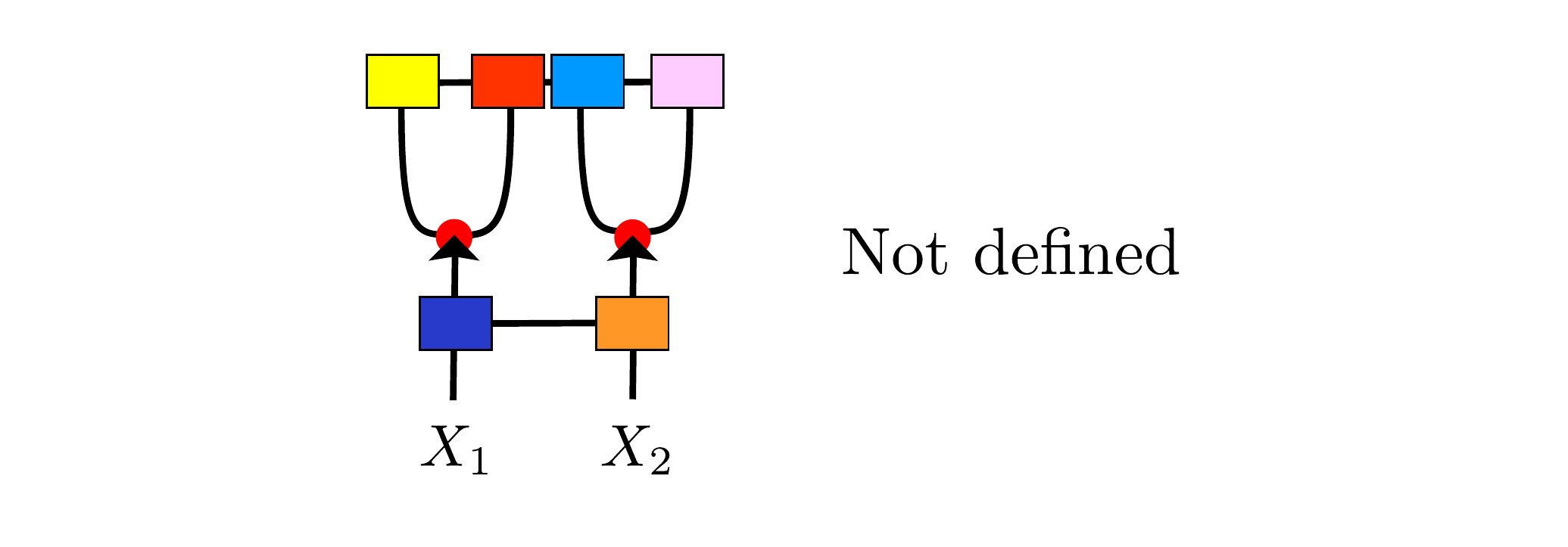}}
\caption{(a) Copy operation of a vector input $A_i$, resulting in a new tensor $B_{ij}=A_i A_j$ (b), (c) and (d) :  Mapping of generalized tensor networks with copy operation to tensor networks with weight sharing (indicated by the same colors of the tensors) and copied input vectors or discrete inputs. (e) This representation is not defined because the copy operation only applies to vectors.}
\label{copyoperation}
\end{figure}

We graphically depict the copy operation through a red dot between edges of the graph and an arrow which marks the incoming edge. The copy operation takes a vector as input, and it outputs two identical copies of it (Fig.~\ref{copyoperation}). This operation is equivalent to having two copies of the input contracted with the rest of the tensor network, and we depict this fact by using the same colors for the same tensors. In the case were the input is a discrete variable, we define the copy operation so that it has the same effect as the copy tensor that simply copies this variable.

More generally, we can use this property to define complex generalized tensor networks where the copy operation is used at different places in the network. The resulting tensor networks are well defined if they can be expressed as a tensor network using several times the same tensor. Generalized tensor networks are therefore tensor networks that use weight sharing between some tensors, and such that several copies of the inputs may be used. In Fig.~\ref{copyoperation} we provide examples of such a mapping to clarify the definition of generalized tensor networks.

\subsubsection{Architectures used in this work}

\begin{figure}[t]
\centering
\subfloat[2D-EPS]{\includegraphics[width = 4cm]{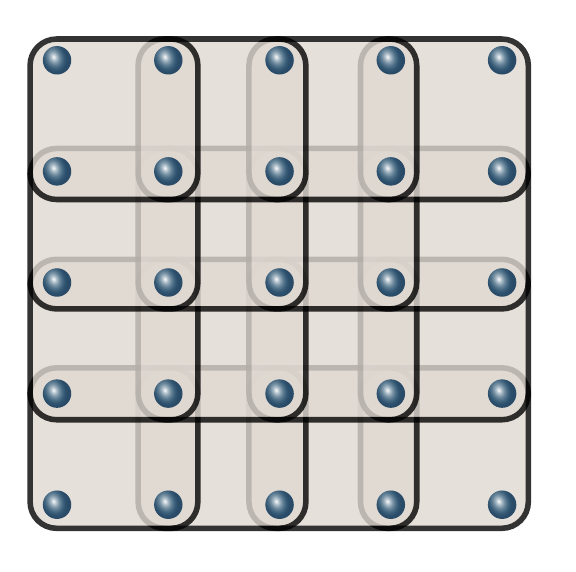}\label{EPS2D}}
\subfloat[2D-SBS]{\includegraphics[width = 3.9cm]{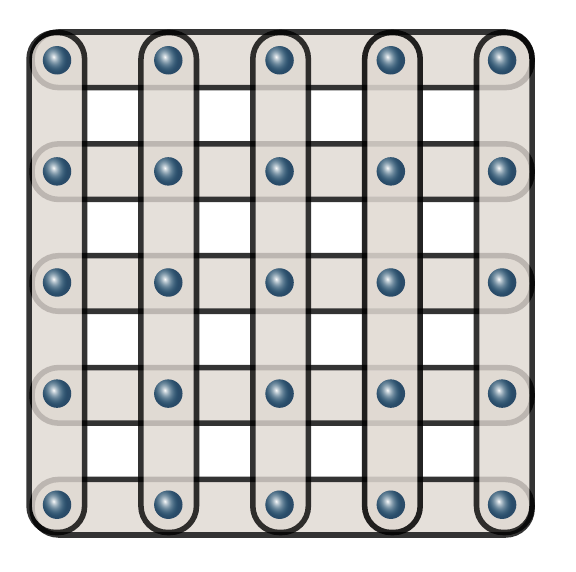}\label{2Dsbs}}\\
\subfloat[Snake-SBS]{\includegraphics[width = 6cm]{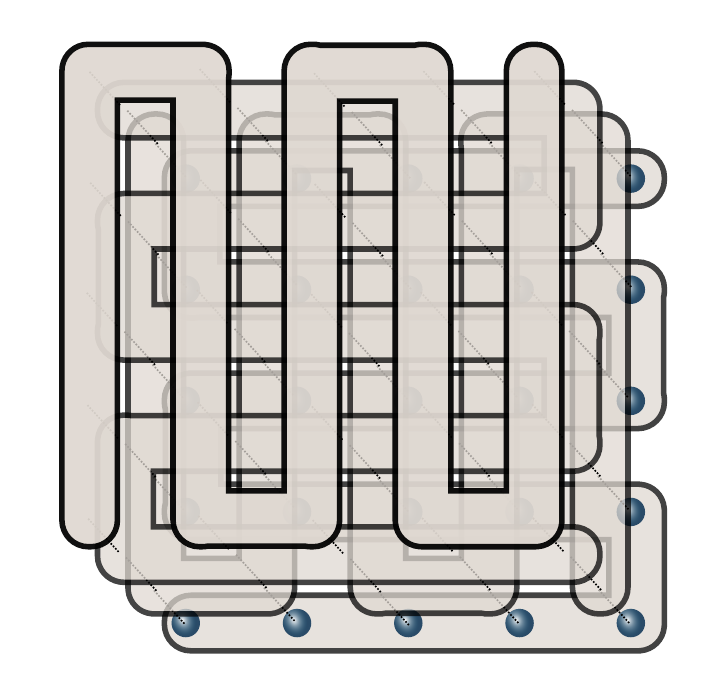}\label{snakesbs}}
\caption{Possible geometries of generalized tensor networks in 2D: (a) EPS. (b) 2D-SBS consisting of horizontal and vertical overlapping strings. (c) Snake-SBS consisting of 4 overlapping strings in a snake pattern.}
\end{figure}

\begin{figure}[t]
\centering
\includegraphics[width = 4cm]{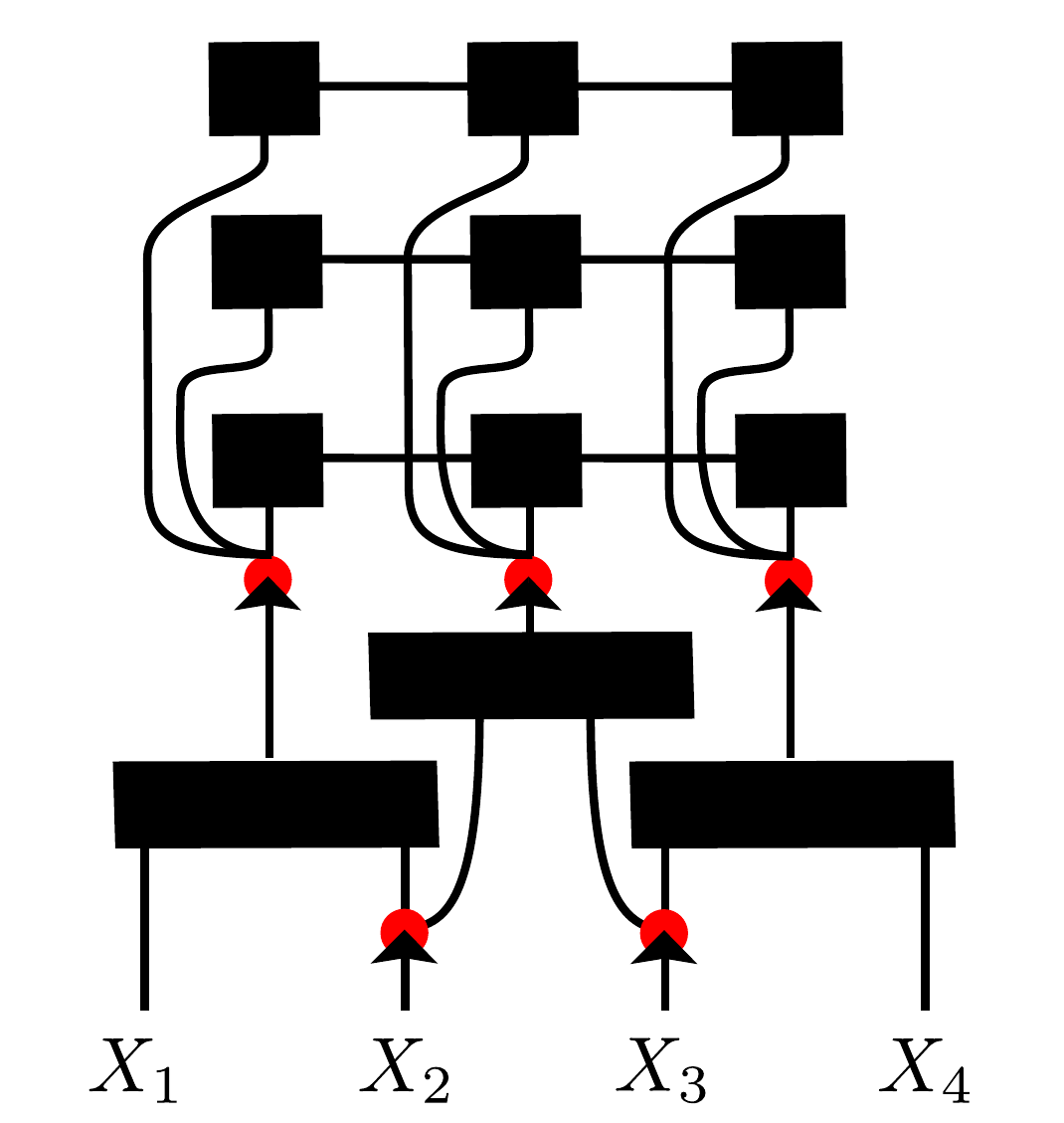}
\caption{EPS-SBS consisting of a first layer of EPS, followed by a copy operation and a second layer of SBS.}
\label{epssbs}
\end{figure}

The previous examples of EPS and SBS can be straightforwardly extended to generalized tensor networks by replacing the copy tensors as inputs by copy operations, so that they can be used with feature vector inputs. More generally, one can think of complex networks built using the copy operation for gluing different networks together. In practice, we will consider the following network architectures, in the case where the inputs have a two-dimensional geometry, such as images:
\begin{itemize}
    \item EPS with $2\times 2$ overlapping plaquettes with weight sharing such that the tensor for each plaquette is the same (Fig.~\ref{EPS2D}).
    \item SBS defined with horizontal and vertical strings covering the 2D lattice (Fig.~\ref{2Dsbs}). We will denote this kind of SBS as 2D-SBS. Correlations along one of the two dimensions can be captured in the corresponding MPS, and more complex correlations are included through the overlap of the different strings.
    \item SBS consisting of 4 strings, each covering the whole lattice in a snake pattern, but in a different order (Fig.~\ref{snakesbs}). We denote these SBS as snake-SBS. They have the advantage, compared to a MPS, that two nearest neighbours variables always appear next to each other in one of the 4 strings, thus rendering the capture of strong local correlations efficient. More complex string geometries could be considered, and the choice of string could be itself learned with a RBM.
    \item EPS with an extra output leg in each plaquette, and such that the outputs are copied and taken as input into a SBS (Fig.~\ref{epssbs}). The input variables are first copied and fed into overlapping clusters parametrized by tensors. The output leg of each of  these tensors is a vector which is copied a few times. Each of these copies can then be contracted with the open legs of a different MPS, forming together a SBS. In 2D, we choose 2x2 overlapping plaquettes in the first layer, and 4 strings forming a snake-SBS in the second layer. In the following we will call this generalized tensor network EPS-SBS.
\end{itemize}
These are simple examples of generalized tensor networks inspired from models invented in quantum physics to capture correlations on 2D lattice systems. More complex networks based on trees or hierarchical designs with more than two layers can also be constructed in the same way. 

We note that there is a relationship between EPS and one layer of a CNN : in the case of discrete inputs, there is a finite number of possible different filters that are applied in a CNN. If we represent the local convolutional operation as a function from discrete inputs, then this function can be represented as a tensor. The fact that convolutional filters are applied with a small stride over an input image corresponds to the overlap of the plaquettes in an EPS, and so EPS correspond to the first layer of a CNN applying all possible filters on discrete inputs.

These generalized tensor networks have the advantage, compared to standard tensor networks, that they can be easily defined in arbitrary dimension and geometry while remaining efficient to contract, as long as the input is fixed. This is in contrast to a 2D tensor network previously introduced in physics known as projected entangled pair states \cite{peps2004}, which is naturally defined in higher dimensions, but cannot be contracted exactly efficiently. In particular, 2D-SBS form a subclass of projected entangled pair states that remains efficient to contract.

Another advantage of generalized tensor networks is that they can represent some functions with fewer parameters than regular tensor networks without copy tensors. Indeed, tensor networks such as MPS or tree tensor networks satisfy a constraint known as area law, which implies that they will require a number of parameters that is exponential in the number of variables to represent functions not satisfying this constraint, while generalized tensor networks such as SBS or RBM can represent some of these functions with a polynomial number of parameter \cite{Deng2017,Levine2018}.

\section{Supervised Learning Algorithm}
\label{section2}

Graphical models are often used in conjunction with unsupervised learning algorithms, since they are designed to represent probability distributions. In particular cases it is possible to compute the normalization $Z$, which gives exact access to the likelihood and makes maximum likelihood estimation tractable. This is possible for graphical models and tensor networks on trees and has led to an algorithm for generative modeling with MPS \cite{Han2017,Glasser2019}. In the more general case, which includes RBM, the normalization $Z$ cannot be computed efficiently. Approximate algorithms relying on Monte Carlo sampling can then be used, such as contrastive divergence \cite{Hinton1985,Hinton2002}. Generalized tensor networks suffer from the same issue, which makes unsupervised learning computationally expensive. Since these networks correspond to quantum states, it might be possible to implement them on a quantum computer and sample from them efficiently. In this work we focus instead on supervised learning, where access to the normalization $Z$ is not necessary. In this section we discuss how RBM can be used for supervised learning, and generalize the corresponding algorithm to tensor networks.

\subsection{Supervised Learning with Restricted Boltzmann Machines}

\begin{figure}[t]
\centering
\subfloat[Discriminative RBM]{\includegraphics[width = 6cm]{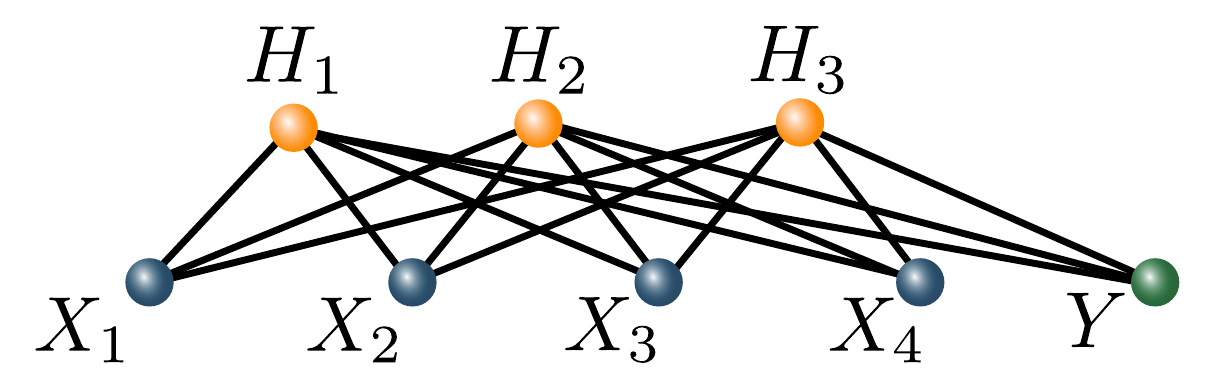}
\label{supervisedRBM}} \\
\subfloat[Discriminative SBS]{\includegraphics[width = 6cm]{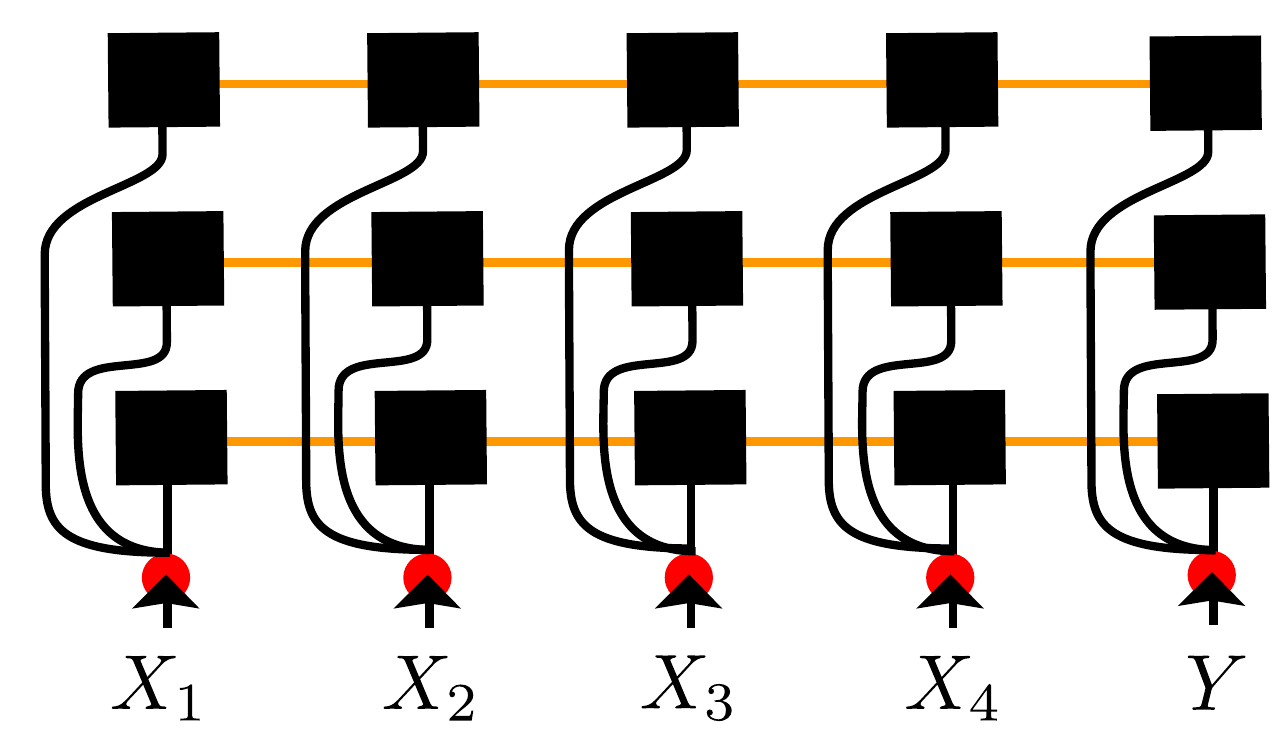} \label{supervisedSBS}} \\
\subfloat[Discriminative EPS]{\includegraphics[width = 6cm]{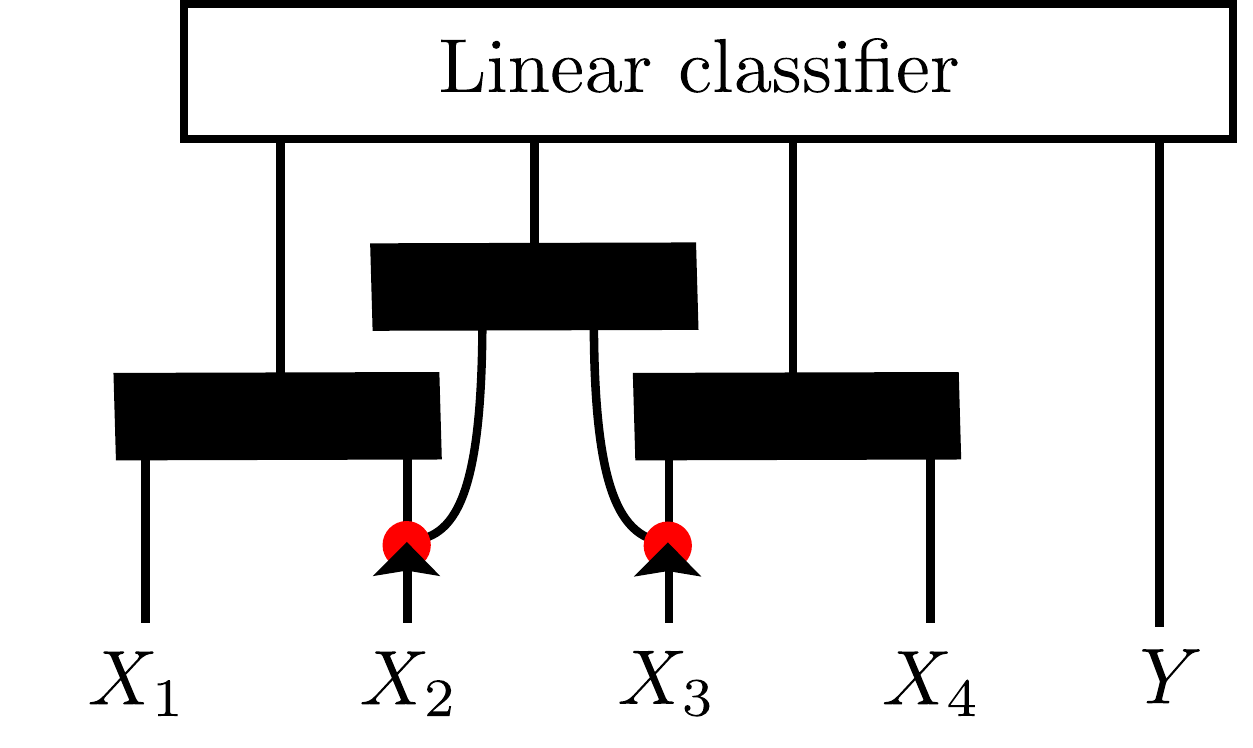} \label{supervisedEPS}}

\caption{(a) A classification RBM turns the label into an additional visible unit. (b) The same procedure can be defined for a SBS by adding a node corresponding to the label, and corresponding tensors which connect it to the rest of the tensor network. (c) Generalized tensor networks can be combined with additional layers of neural networks. For example an EPS output is a tensor that can be combined with a linear classifier.}
\end{figure}

We first review how RBM can be used to perform supervised learning, in a classification setting \cite{Larochelle2008,Larochelle2012}. Given labelled training data $\mathcal{D}=\{(\mathbf{x}_i,y_i)\}$, where the $y_i$ take discrete values corresponding to different classes, a RBM can be used to approximate the joint probability distribution of the variables and labels:
\begin{align}
p(\mathbf{x},y)=\frac{1}{Z}  \sum_{\mathbf{h}} e^{\mathcal{H}(\mathbf{x},\mathbf{h},y)} 
\end{align}
In such a model, the label is seen as an additional visible variable (Fig.~\ref{supervisedRBM}), possibly encoded in a one-hot representation to use only binary units. Training such a generative model can be done by maximizing the log-likelihood of the data. Since the likelihood is intractable, because the partition function $Z$ cannot be efficiently computed, such a training can be done through approximate algorithms such as contrastive divergence. In supervised learning, one is interested in computing the conditional distribution
\begin{align}
p(y|\mathbf{x})=\frac{p(\mathbf{x},y)}{\sum_{y_j} p(\mathbf{x},y_j)},
\end{align}
which can be computed exactly when the number of classes is small enough, since the two partition functions cancel. The label predicted by the model for new data $\mathbf{x}_i$ is the label maximizing  $p(y|\mathbf{x}_i)$. Since one is ultimately interested in classification performance, it can be advantageous to directly optimize $p(y_i|\mathbf{x_i})$, which leads to a cost function to minimize
\begin{align}
\mathcal{L_{\text{discriminative}}}=-\sum_{i=1}^{|D|} \log p(y_i|\mathbf{x}_i),
\end{align}
whose gradient can be computed analytically. A RBM can moreover be trained in a semi-supervised way, by using a combination of discriminative and generative training.

\subsection{Supervised Learning with Generalized Tensor Networks}

To generalize the discriminative training of RBM to generalized tensor networks, we would like to approximate the joint probability distribution of the variables and labels as a tensor network:
\begin{align}
p(\mathbf{x},y)\propto\text{GTN}(\mathbf{x},y),
\end{align}
where $\text{GTN}(\mathbf{x},y)$ is the function resulting of the contraction of a generalized tensor network with the inputs features and with the discrete label variable. More specifically, in the case of discrete variables the variables $\mathbf{x}$ and $y$ fix the values of the open legs of the generalized tensor network, and the whole network is contracted, resulting in a scalar. In the case of input variables that are feature vectors, each vector is contracted with the corresponding open leg of the tensor network. In order for the cost function to be well defined, the result of the network contraction should be positive. This can be done by ensuring that the tensor elements are positive, as in a RBM or graphical model, or by choosing instead $p(\mathbf{x},y)\propto\left(\text{GTN}(\mathbf{x},y)\right)^2$, which corresponds to a Born machine \cite{Cheng2017,Liu2018b,Glasser2019}, or $p(\mathbf{x},y)\propto e^{\text{GTN}(\mathbf{x},y)}$. In the following we will adopt this last choice, for which training is found to be more efficient.The label is now seen as the index of one tensor. Since it is discrete, there is no need to use a one-hot representation and one can simply enlarge the dimension of the leg of a tensor to accommodate for the number of possible classes (Fig.~\ref{supervisedSBS}). We then define, by analogy with the graphical model case,
\begin{align}
p(y_k|\mathbf{x_i})=\frac{e^{\text{GTN}(\mathbf{x_i},y_k)}}{\sum_{y_j} e^{\text{GTN}(\mathbf{x_i},y_j)}},
\end{align}
which can be seen as a softmax function applied to the different outputs of the tensor network contraction, and the cost function is again chosen to be a cross-entropy loss
\begin{align}
\mathcal{L_{\text{discriminative}}}=-\sum_{i=1}^{|D|} \log p(y_i|\mathbf{x_i}).
\end{align}
The gradient of the cost function can then be expressed  as a function of $\text{GTN}(\mathbf{x_i},y_j)$ and its derivatives, since
\begin{align}
\frac{\partial \log p(y_i|\mathbf{x_i})}{\partial w}=\frac{\partial \text{GTN}(\mathbf{x_i},y_i)}{\partial w}-\sum_{y_j} p(y_j|\mathbf{x_i}) \frac{\partial \text{GTN}(\mathbf{x_i},y_j)}{\partial w}.
\end{align}
$\text{GTN}(\mathbf{x_i},y_j)$ can be computed exactly by fixing the value of the input units and labels and contracting the network. We note that in general we can contract the whole network without the labels, and perform the contraction for the different labels as a last step. Contraction of the whole network for different labels thus only adds a small cost (which depends on the shape of the network) to the contraction of the network without labels. We observe that from the point of view of supervised learning there is no essential difference between SBS and MPS in terms of the optimization algorithms: the cost of optimizing a SBS is only a constant factor (the number of strings) more expensive than for a MPS. This is unlike in quantum physics where Monte Carlo sampling is necessary to optimize a SBS. To further regularize the tensor network, we randomly drop tensor elements to $0$ with probability $\delta$ during training, following the procedure in \cite{Novikov2016}.

So far we have constructed tensor networks which, when an input and a label is given, have no open legs. We can also construct networks with open legs and use tensor networks in combination with other machine learning techniques. In this case the tensor network maps the input to a tensor which can for example be used as input in a neural network. In the case of EPS where each tensor over overlapping plaquettes has an open leg, an input is mapped to a tensor with an extra dimension as output. The simplest way to combine EPS with other neural networks is to place a linear classifier on top of the EPS (Fig.~\ref{supervisedEPS}). The backpropagation algorithm used to compute derivatives of the neural network is in this case combined with the algorithm for computing derivatives of a tensor network, and the joint network can be optimized using stochastic gradient descent. More complex combinations of tensor networks and neural networks may be defined in the same way.

\section{Learning Feature Vectors of Data}
\label{section4}

In this section we explore several strategies that can be used to deal with data that is not discrete. We suggest to learn relevant tensor features as part of the tensor network and discuss how tensor features can also be learned as part of a deep learning architectures which combines a neural network extracting features with a tensor network.\\

\begin{figure}[t]
 \centering
 \subfloat[]{\includegraphics[width = 4cm]{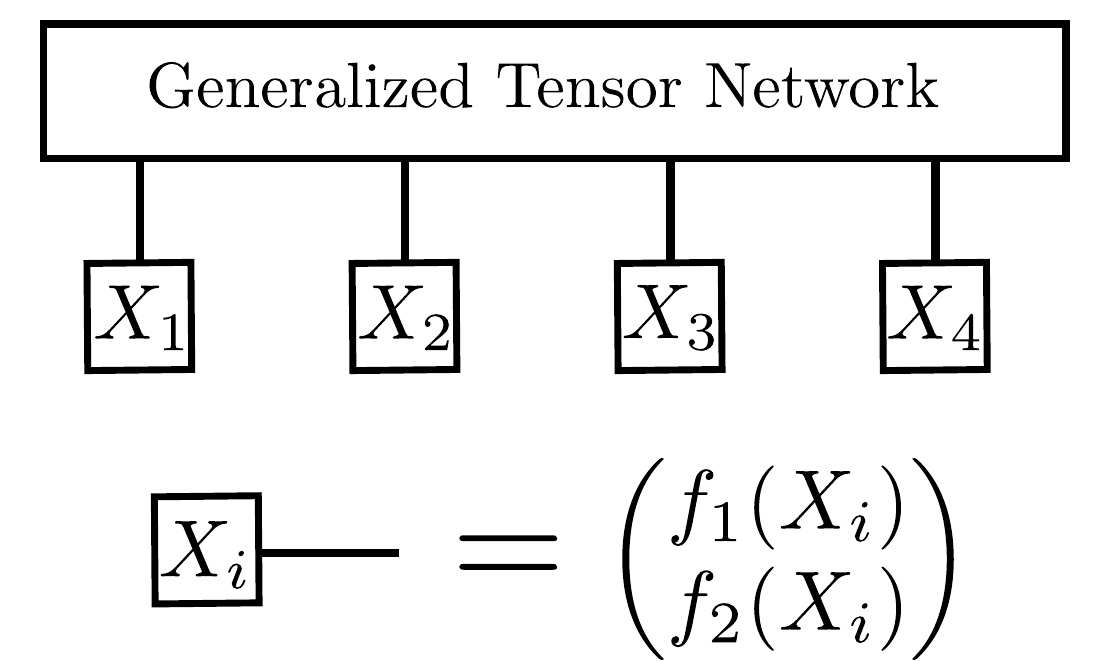}\label{featuresinput}}
 \subfloat[]{\includegraphics[width = 4cm]{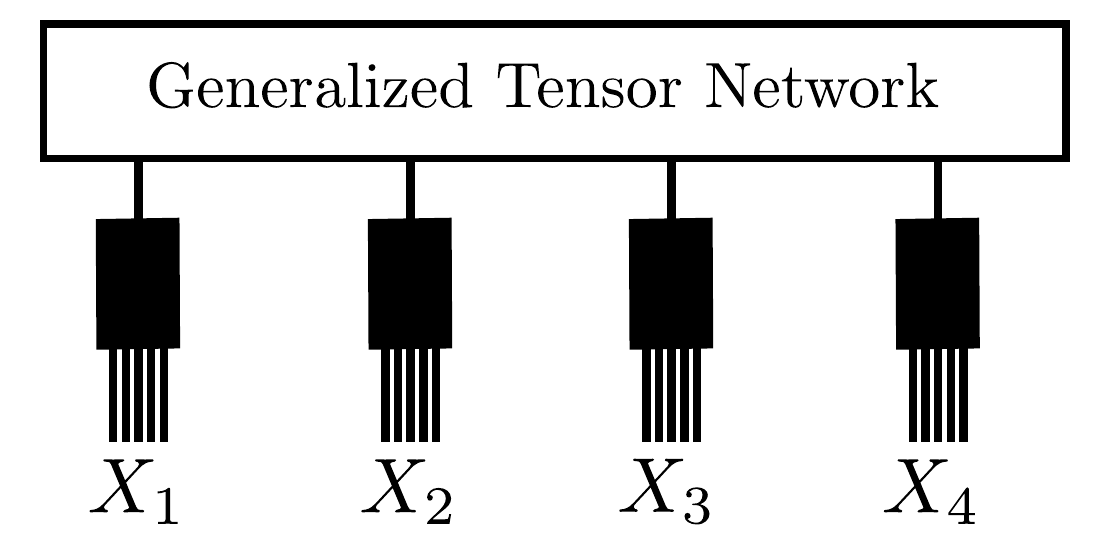}\label{featurelearning}}
 \caption{(a) Real inputs $X_i$ are mapped to a feature vector (here with length two). This vector can then be used as input to a generalized tensor network by contracting it with the open legs of the generalized tensor network.
 (b) Feature tensors can compress the discretized representation of the inputs $X_i$ to a smaller dimensional space. These tensors can share weights and can be learned as part of the tensor network.}
 \end{figure}

A naive way of applying tensor networks with real data would be to discretize data or use its binary representation. This is not a suitable approach, because that would amount to  dramatically increasing the size of the data, rendering learning very slow, and it would also lead to large tensor networks prone to overfitting. Another approach, as suggested by \cite{Stoudenmire2016} and \cite{Novikov2016}, is to map the real data to a higher dimensional feature space. Each variable is first independently mapped to a vector of length (at least) two in order to be contracted with the open legs of the tensor network (Fig.~\ref{featuresinput}). Choices of feature maps that have been used include
\begin{align}
x\rightarrow \begin{pmatrix}
1 \\ x
\end{pmatrix}\ \text{or} \ 
\begin{pmatrix}
\cos(\frac{\pi}{2} x) \\ \sin(\frac{\pi}{2} x)
\end{pmatrix},
\end{align}
and generalizations to higher dimensions. A choice which is suitable with our algorithm, assuming that the data is normalized between $0$ and $1$, is to use 
\begin{align}
x\rightarrow \begin{pmatrix}
\cos^2(\frac{\pi}{2} x) \\ \sin^2(\frac{\pi}{2} x)
\end{pmatrix},\label{featurechoice}
\end{align}
because this ensures that the vectors are positive and the normalization prevents numerical instabilities.

\begin{figure}[t]
 \centering
 \subfloat[]{\includegraphics[width = 4cm]{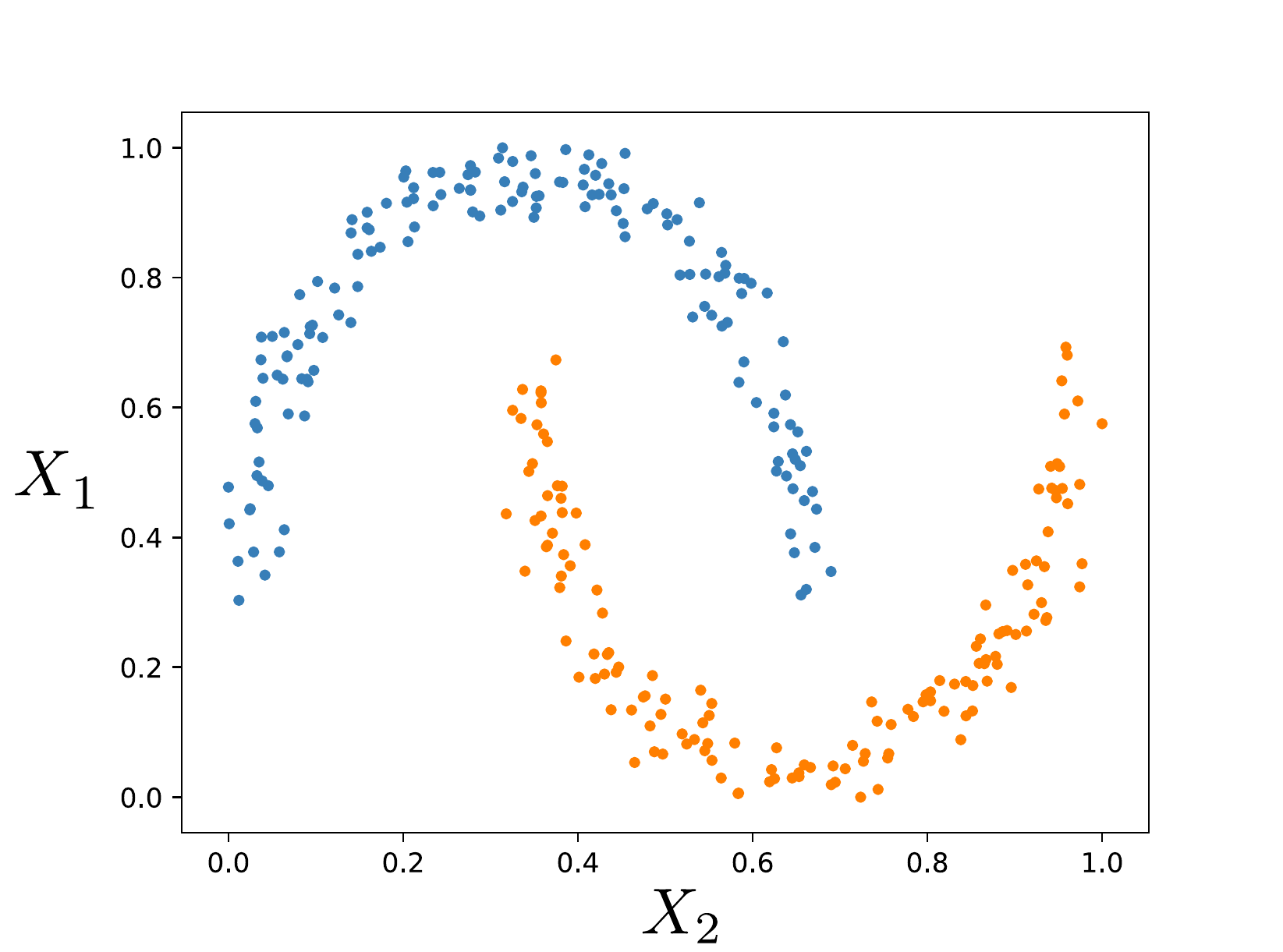}\label{artificialdataset}}
 \subfloat[]{\includegraphics[width = 4.33cm]{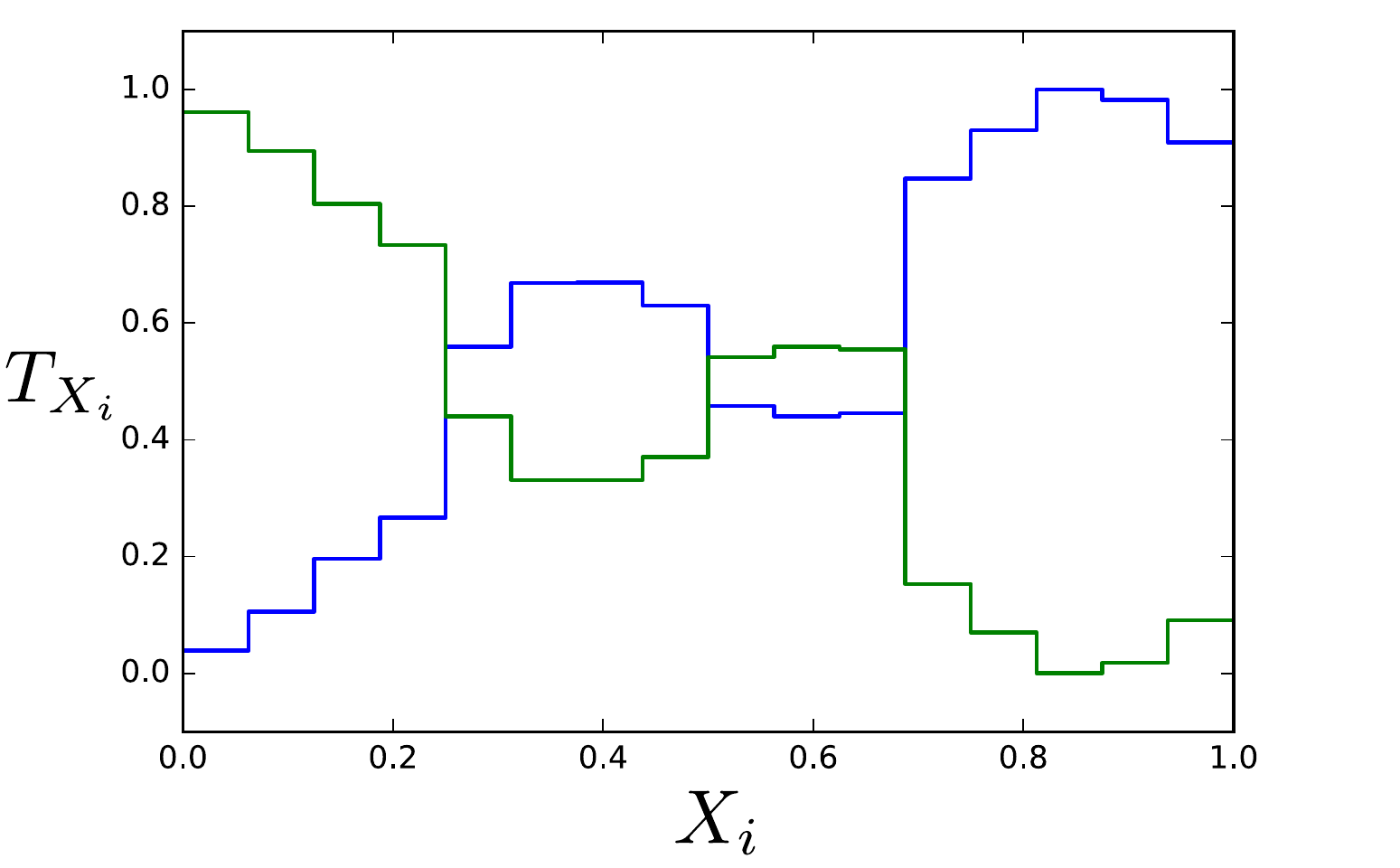}\label{artificialdatasetfeatures}}
 \caption{
 (a) Dataset with two features $X_1$ and $X_2$ and two classes (depicted in different colors) that cannot be learned by a MPS of bond dimension 2 with the feature choice in \eqref{featurechoice}. (b) Two normalized features learned by a tensor while classifying the previous data set with a MPS of bond dimension 2. The features have been discretized in 16 intervals. Using this choice of features the MPS can classify the data set perfectly.}
 \end{figure}

These choices however put severe limitations to the functions that can be learned. Indeed, the data set with just two variables presented in Fig.~\ref{artificialdataset} cannot be separated by a MPS of bond dimension 2 with one of these feature choices, since the boundary decision will be a polynomial of degree two of the features. Nevertheless, a different feature choice could distinguish the two classes, even with bond dimension 2. We therefore suggest to learn the appropriate features as part of the learning algorithm. This can be done by parametrizing the feature functions and learning them at the same time as the rest of the network. To be able to use a purely tensor network algorithm, we can parametrize these functions using a tensor network. In the simplest case, we discretize the real data and use a tensor to compress the large dimensional input into a smaller dimensional vector of suitable length. This tensor can be learned as part of the whole tensor network and prevents the size of the rest of the tensor network to increase when the discretization size changes. The feature tensor can be the same for all variables, for example image pixels, but can be different in the case where the variables are of different nature. Using this procedure, a MPS of bond dimension 2 is able to get perfect accuracy on the data set presented in Fig.~\ref{artificialdataset}. The two features that the network has learned are presented in Fig.~\ref{artificialdatasetfeatures}. We note that starting from random features on more complex data sets makes learning difficult, but the feature tensor can be pretrained using a linear classifier, before being trained with the rest of the network.

\begin{figure}[t]
 \centering
 \subfloat[]{\includegraphics[width = 4cm]{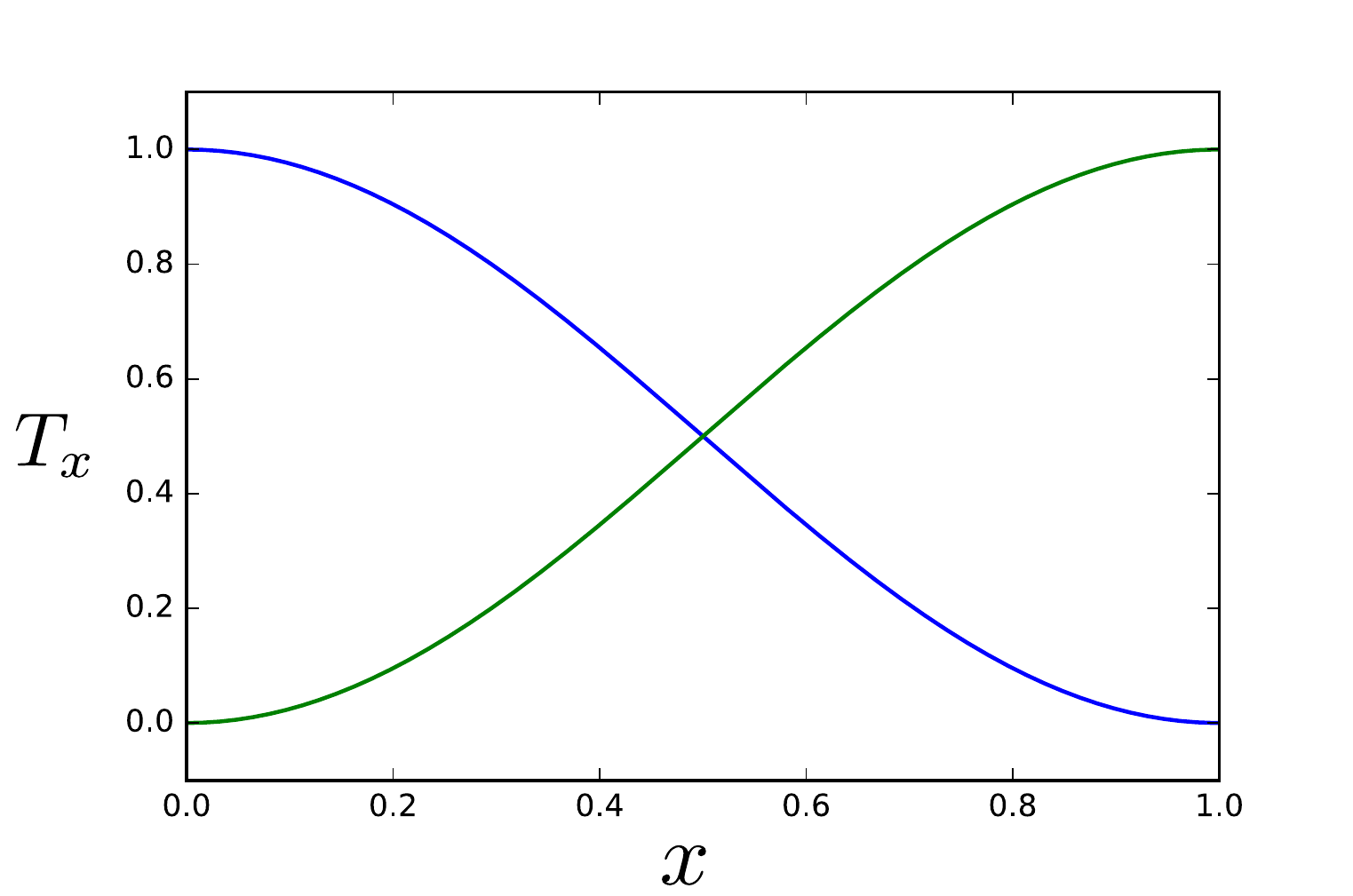}\label{featureschoiceplot}}
 \subfloat[]{\includegraphics[width = 4cm]{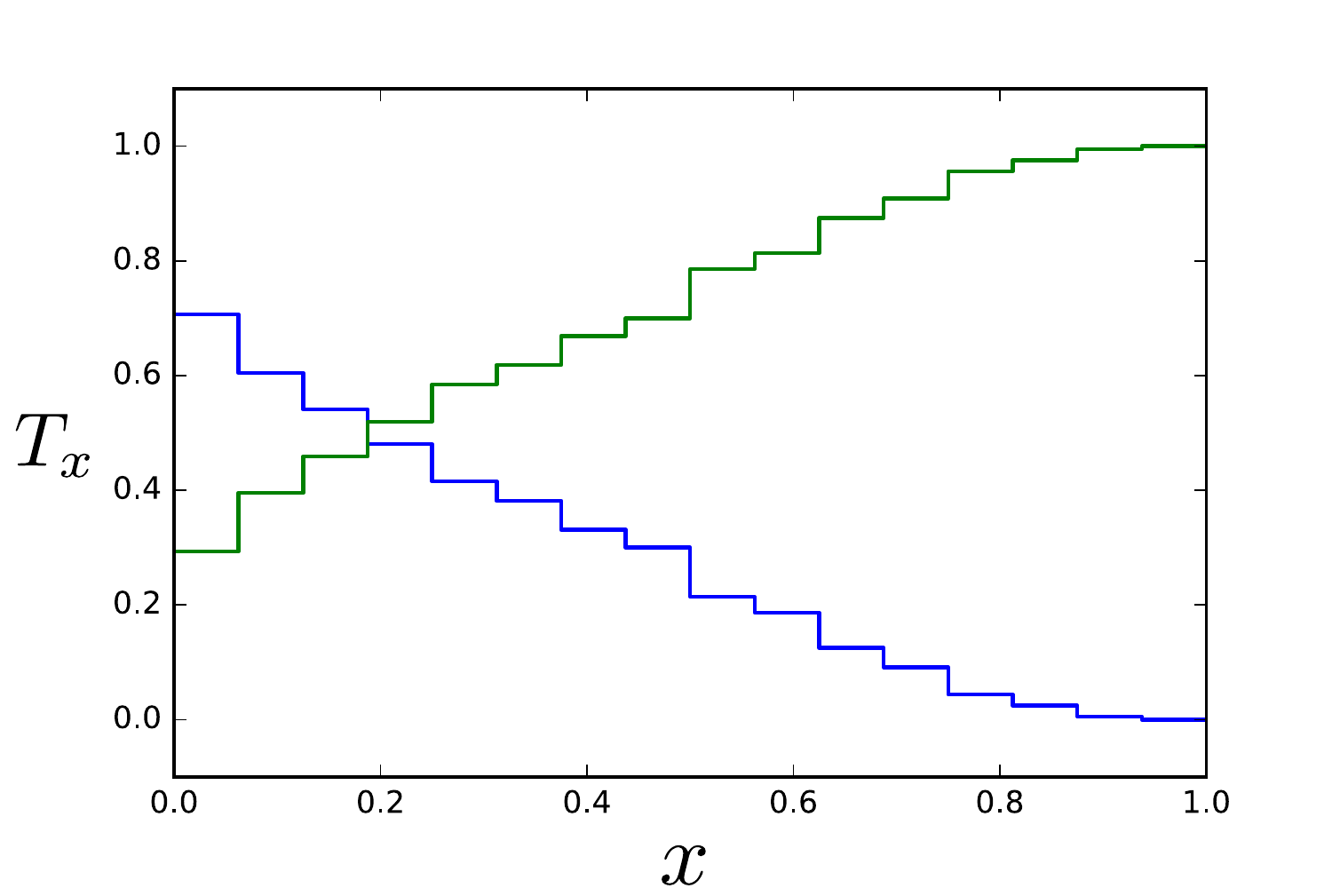}\label{featureslearnedMNIST}}
 \caption{(a) Choice of two features in \eqref{featurechoice} for an input taking real values between 0 and 1.
 (b) Two normalized features learned by a tensor with output dimension 2 combined with a snake-SBS classifying the MNIST data set. The input features $x$ are the greyscale value of pixels, normalized between 0 and 1 and discretized in 16 intervals.}
 \end{figure}

In comparison, we also show in Fig.~\ref{featureslearnedMNIST} the features learned while classifying MNIST with greyscale pixels and a snake-SBS (see Section \ref{section5}). These features are not very different from the choice in \ref{featurechoice}, and we could not distinguish performance with this choice or with learned features on this data set. We expect however that this procedure will be necessary for more complex data sets which are not easily approximated by a binary function. Moreover the size of the feature vector provides a regularization of the model, and higher sizes might be necessary for more complex data sets. More generally this tensor could be itself represented with a small tensor network, to prevent the number of parameters to increase too much with a very small discretization interval. It is interesting to note that the features learned in our examples are almost continuous even if we use smaller discretization intervals. This means that two real inputs that are close to each other will lead to the same predictions by the network, a property which is in general not true if we simply discretize the inputs and use a larger tensor network. Our approach of learning the features as part of the tensor network may be especially relevant in the context of quantum machine learning, where the tensor network is replaced by a quantum circuit and it might be suitable to have the full network as part of the same quantum machine learning architecture.

\begin{figure}[t]
 \centering
 \includegraphics[width = 8cm]{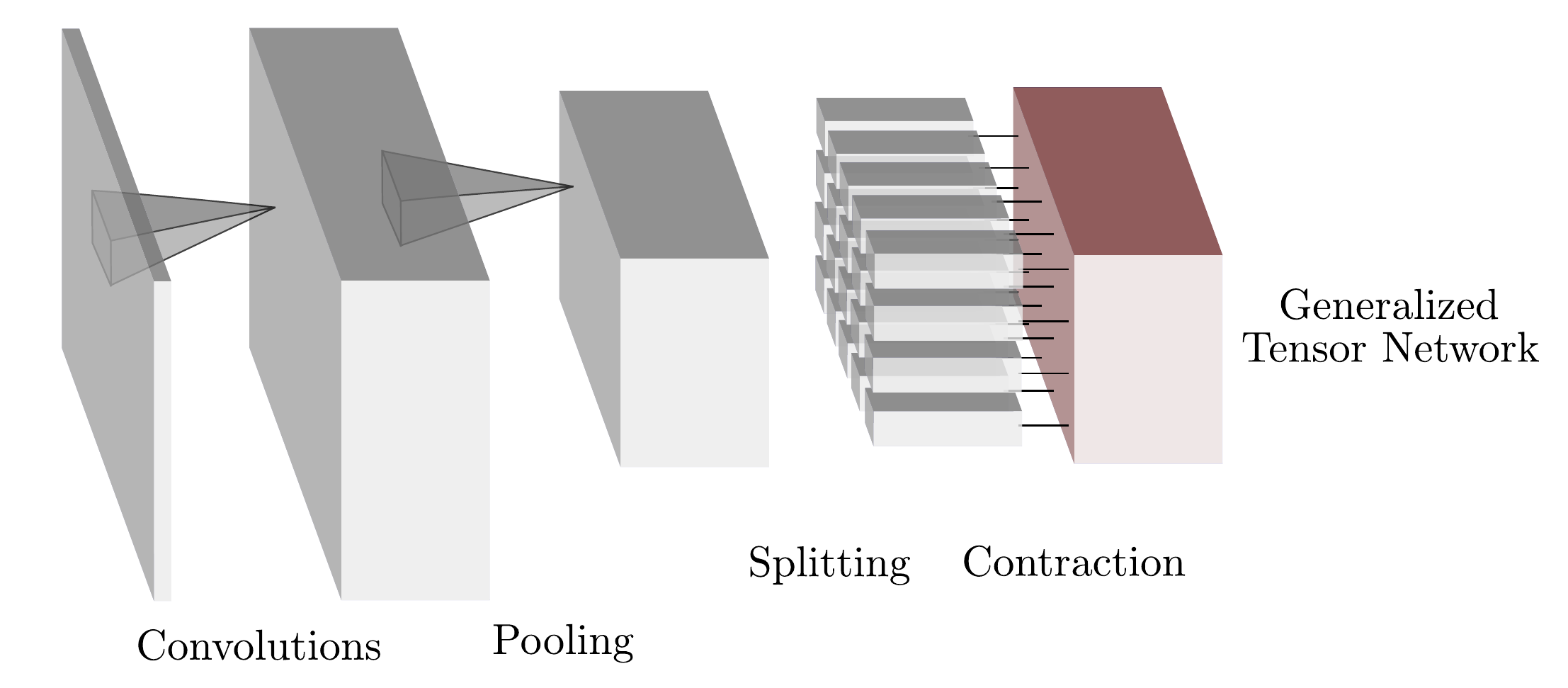}
 \caption{Using Convolutional Neural Networks as feature vector extractors from real data: the output of the CNN is seen as an image with a third dimension collecting the different features. For each pixel of this image, the vector of features is contracted with the open legs of a tensor network.}
 \label{cnn}
 \end{figure}

As an alternative way of choosing the features, we can combine the feature choice with other machine learning techniques. If the input data represents images, it is a natural choice to use Convolutional Neural Networks as feature extractors, since these have been highly successful for image classification. CNN consist in convolution filters, which use convolutional kernels to transform an image into a set of filtered images, and pooling layers which downsize the images (Fig.~\ref{cnn}). The resulting features preserve a form of locality. Therefore it is natural to consider the vector of applied filters associated with each location in the image as a feature vector that can be used in conjunction with generalized tensor networks. The CNN and the tensor network can be trained together, since the derivatives of the tensor network can be used in the backpropagation algorithm which computes the gradient of the cost function.

\section{Numerical Experiments}
\label{section5}

We test the generalized tensor network approach on the task of image classification, where a natural two-dimensional geometry that can be reflected in the architecture of the tensor network is present, as well as on the task of urban sound recognition, where the time dimension provides a one-dimensional geometry.

\subsection{Image Classification}

We first consider the MNIST data set \cite{MNISTreference}, which consists of $28 \times 28$ greyscale images of digits. There are 10 classes and we adopt a multiclass classification procedure in which one tensor of the tensor network is parametrized by the ten possible labels. The original training set is split into training and validation sets of 55000 and 5000 examples and the performance of the different models is evaluated on the test set of 10000 examples. We consider the following generalized tensor networks: a snake-SBS with 4 strings (Fig.~\ref{snakesbs}), a 2D-SBS (Fig.~\ref{2Dsbs}), an EPS with a $2\times 2$ translational-invariant plaquette combined with a linear classifier, (Fig.~\ref{supervisedEPS}), an EPS-SBS with translational-invariant plaquette combined with a snake-SBS (Fig.~\ref{epssbs}) and a CNN-snake-SBS which uses a 1-layer CNN as input features (Fig.~\ref{cnn}). The CNN considered here uses a convolutional layer applying 6 $5\times 5$ filters (stride 1) with ReLU activation function and a pooling layer performing max pooling with a $2\times 2$ filter. All other networks use the choice of features presented in \eqref{featurechoice} and the greyscale values are normalized between 0 and 1. We compare the performance of these networks with a MPS \cite{Stoudenmire2016} and a RBM (the number of hidden units of 250, 500, 750 or 1000 is taken as a hyperparameter). All networks use a batch size of $20$ examples and hyperparameters such as the learning rate $\alpha$ and the regularization rate $\delta$ are determined through a grid search while evaluating the performance on the validation set. Best performance is typically achieved with $\alpha=10^{-4}$, $\delta=0.95$ and a hundred epochs of training. 

 \begin{figure}[t]
 \centering
 \subfloat[]{\includegraphics[width = 4cm]{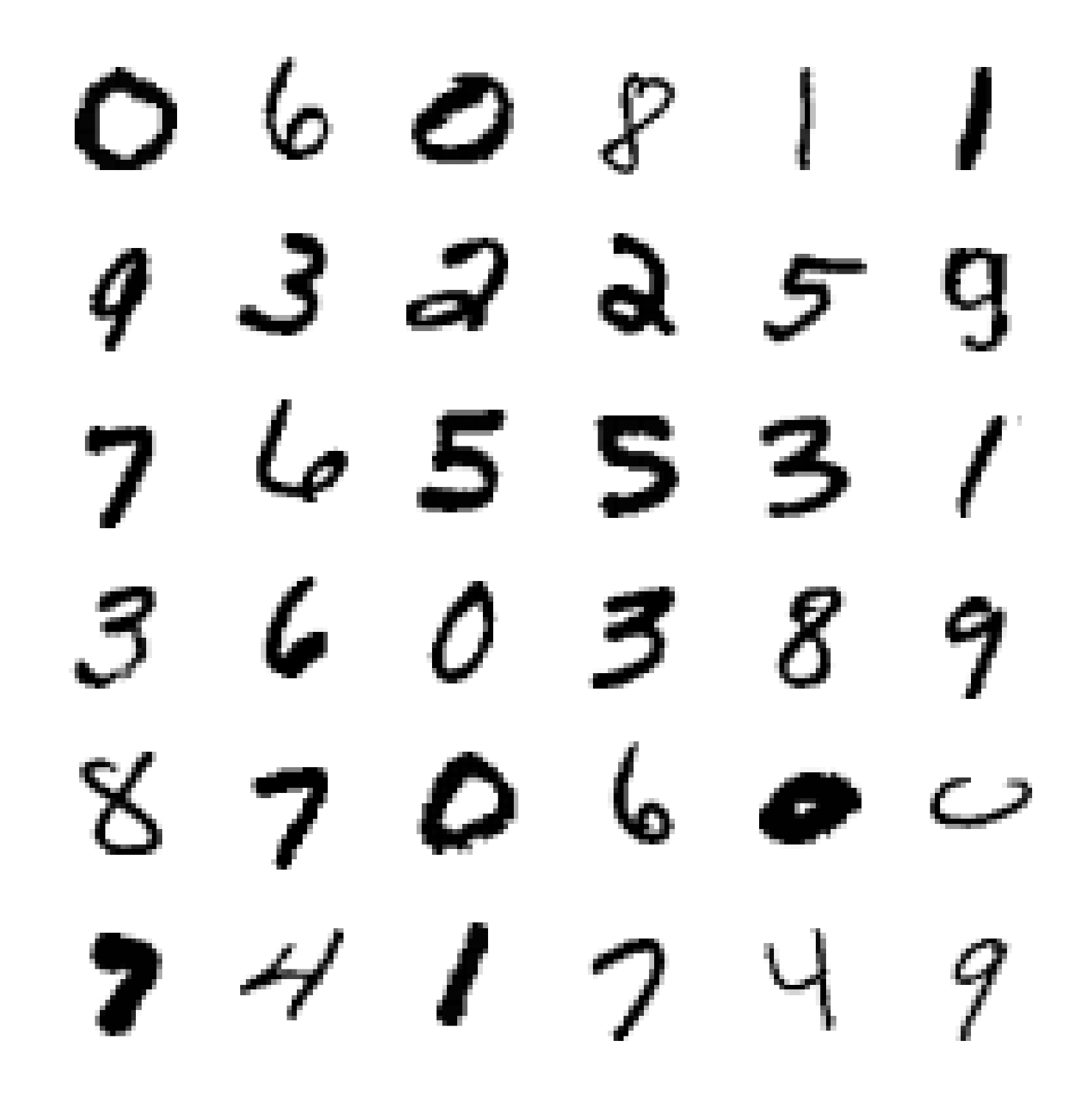}\label{mnist}}
 \subfloat[]{\includegraphics[width = 4cm]{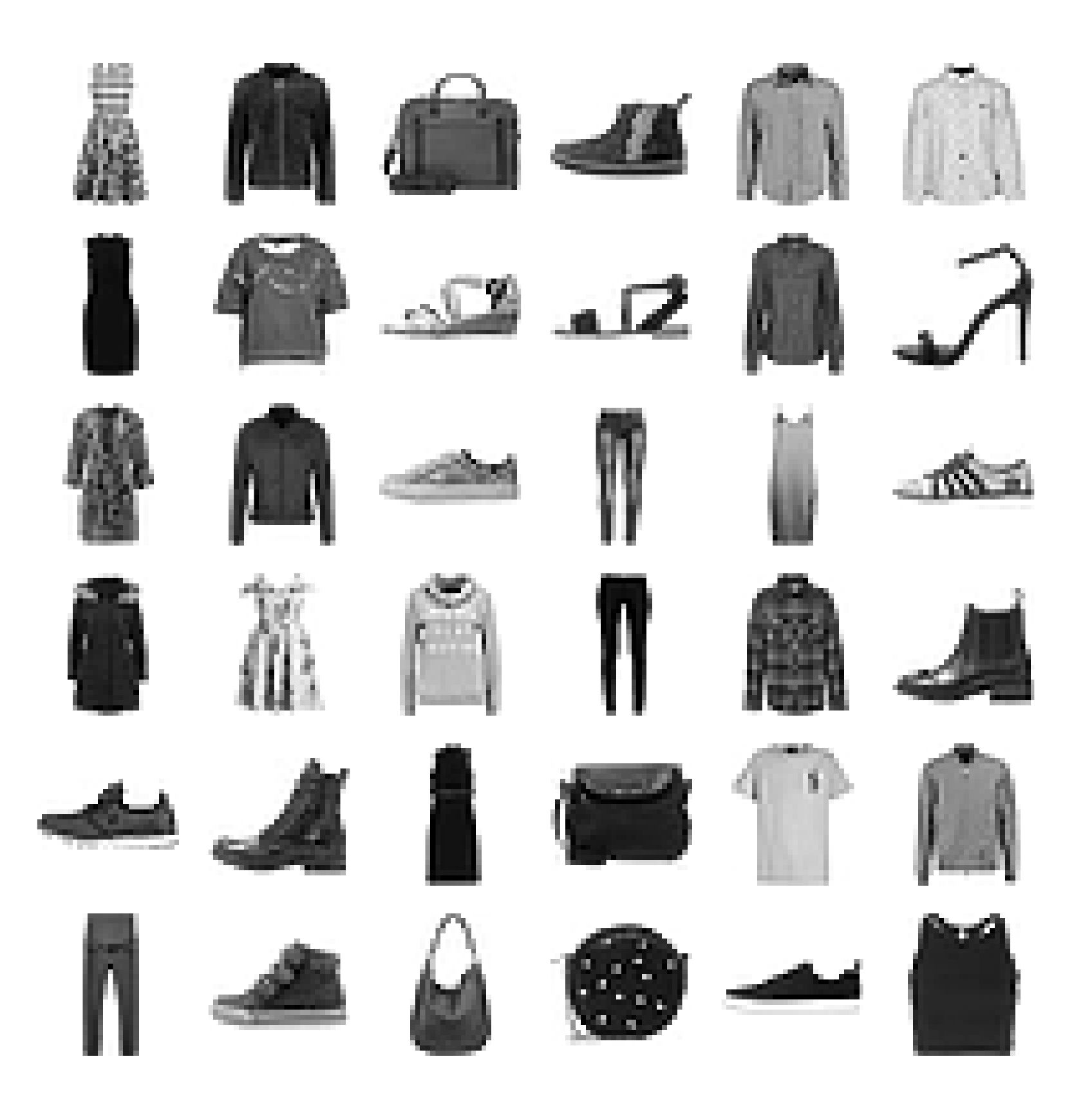}\label{fashionmnist}}
 \caption{Examples of images from the MNIST (a) and fashion MNIST (b) data set.}
 \end{figure}

 \begin{figure}[t]
 \centering
 \includegraphics[width = 8cm]{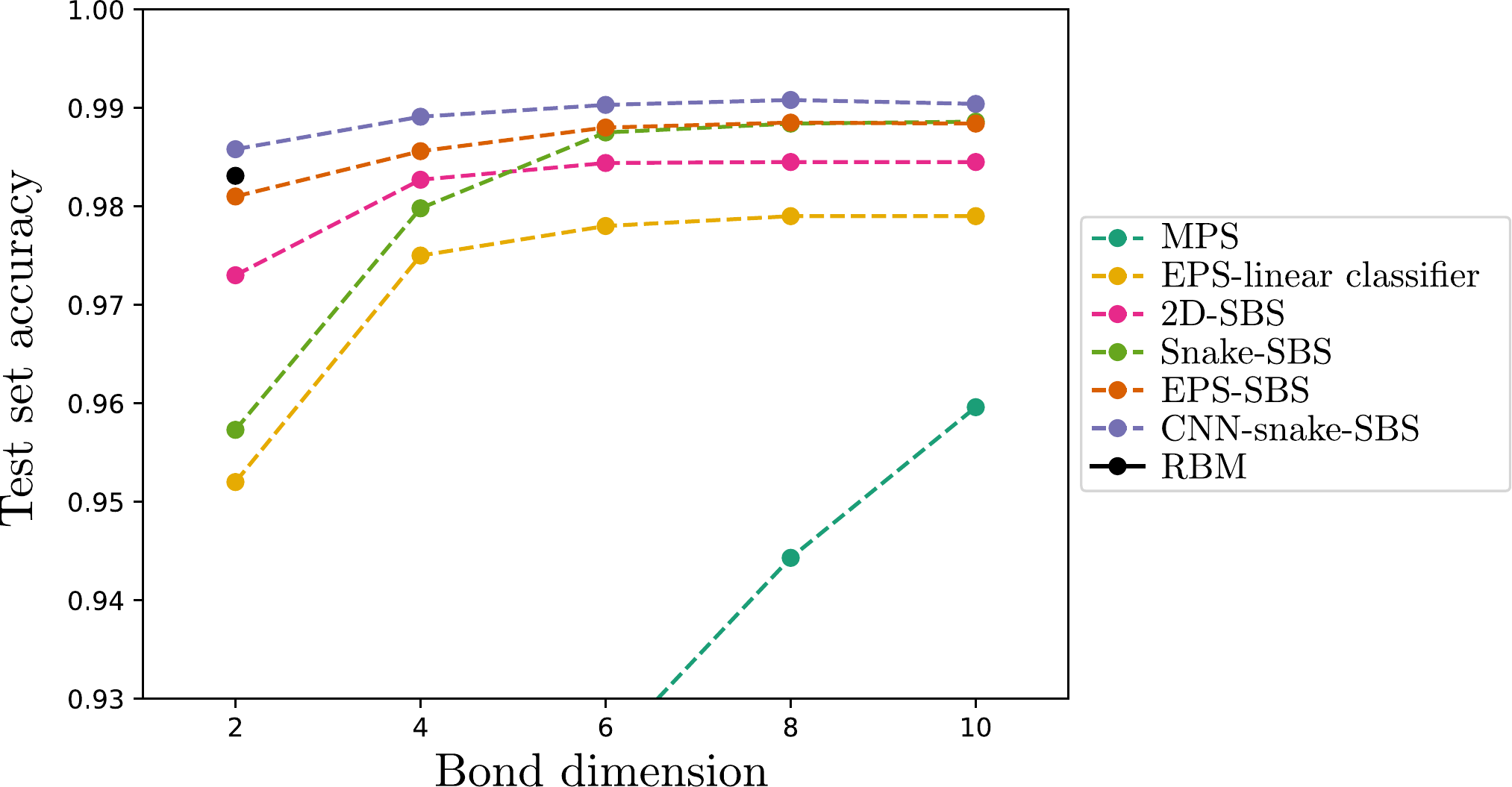}
 \caption{Test set accuracy of different generalized tensor networks on the MNIST data set.}
 \label{mnistresults}
 \end{figure}

The test set accuracy, presented in Fig.~\ref{mnistresults}, shows that even with a very small bond dimension generalized tensor network are able to accurately classify the data set. Their performance is significantly better than that of a tree tensor network \cite{Liu2017} or a MPS trained in frequency space\cite{Liu2018}, and while a MPS can also achieve $99.03\%$ accuracy with a bond dimension of $120$ \cite{Stoudenmire2016}, the cost of optimizing very large tensors has prohibited the use of this method for larger problems so far. The snake-SBS with bond dimension larger than 6 has also better performance than a RBM. Since the snake-SBS provides an interpolation between RBM and MPS, the choice of number of strings and geometry can be seen as additional parameters which could be tuned further to improve over the performance of both methods. All networks have a training set accuracy very close to $100\%$ when the bond dimension is larger than 6, and we expect that better regularization techniques or network architectures have to be developed to significantly increase the test set performances obtained here. We also optimized a snake-SBS with positive elements (by parametrizing each element in a tensor as the exponential of the new parameters), which is a graphical model. Using the same algorithm, we were not able to achieve better performance than $93\%$ classification accuracy with bond dimensions up to $10$. This shows that while having a structure closely related to graphical models, tensor networks may provide different advantages.

\begin{table}[h]
\centering
\begin{tabular}{c c} 
 \hline  \hline
 Method & Accuracy \\
 \hline
Support Vector Machine & $84.1\%$ \\
EPS + linear classifier  & $86.3\%$ \\
Multilayer perceptron & $87.7\%$ \\
EPS-SBS & $88.6\%$ \\
Snake-SBS & $89.2\%$ \\
AlexNet & $89.9\%$ \\
CNN-snake-SBS & $92.3\%$ \\
GoogLeNet & $93.7\%$ \\
 \hline  \hline
\end{tabular}
\caption{Test set accuracy of generalized tensor networks and other approaches\cite{fashionmnistreference} on the fashion MNIST data set.}
\label{tablefashionmnist}
\end{table}

We then turn to the fashion MNIST data set \cite{fashionmnistreference}, consisting of $28 \times 28$ greyscale images of clothes. While having the same size as the original MNIST data set, it is significantly harder to classify. We report the best accuracy obtained with different generalized tensor networks with bond dimension up to 10 in Table~\ref{tablefashionmnist}. It is found that these networks, while not state-of-the-art, are competitive with other approaches such as Support Vector Machines, AlexNet and GoogLeNet Convolutional Neural Networks or a multilayer perceptron neural network, which is encouraging considering the potential improvements in terms of network architecture or training algorithms.

\subsection{Environmental Sound Classification}

\begin{figure}[t]
 \centering
 \includegraphics[width = 7cm]{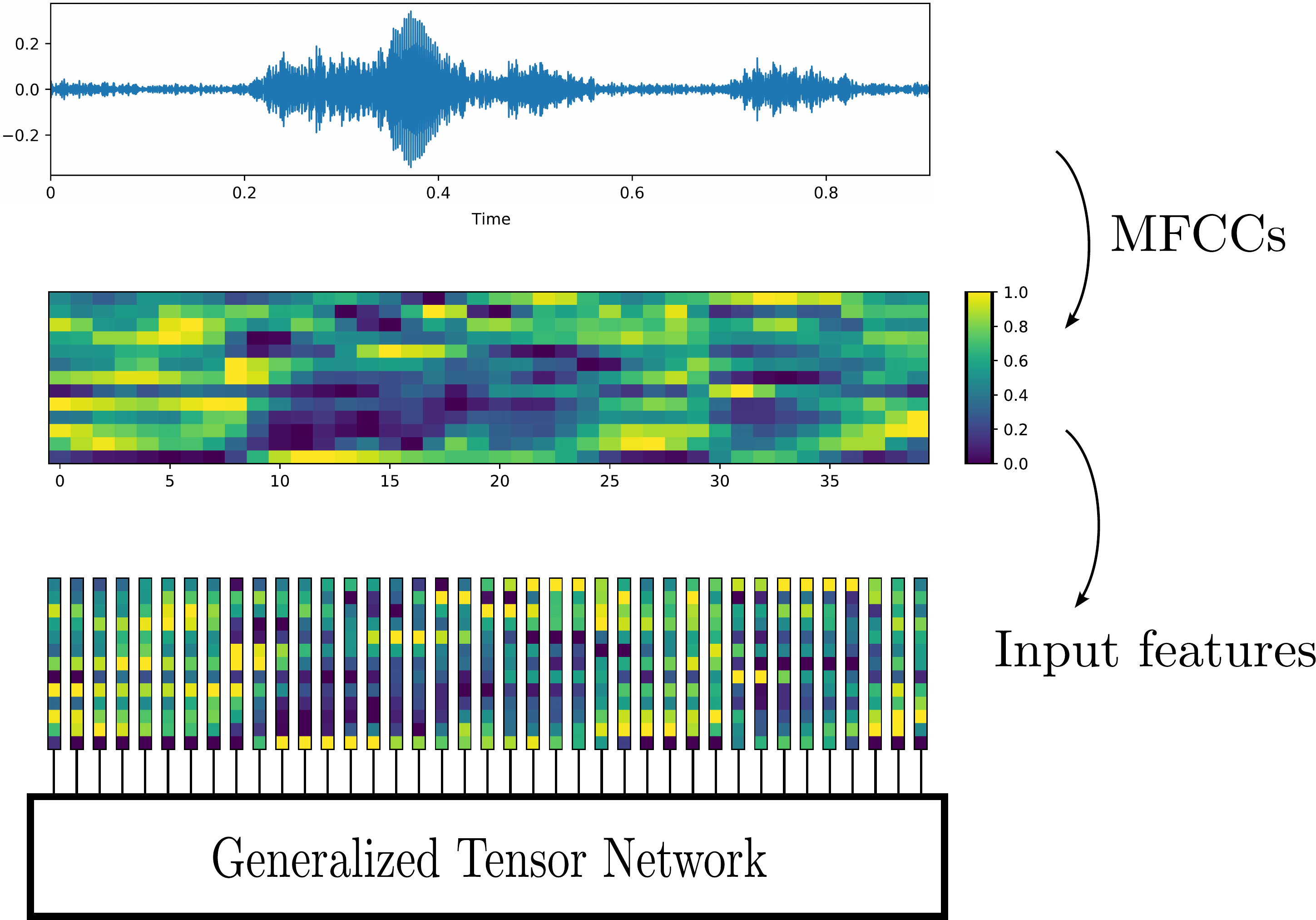}
 \caption{From the raw audio signal, Mel-frequency cepstral coefficients (MFCCs) are extracted over short overlapping windows, resulting in a sequence of high dimensional vectors. These vectors are taken as input to a generalized tensor network.}
 \label{sound}
 \end{figure}

\begin{figure}[t]
 \centering
 \includegraphics[width = 8cm]{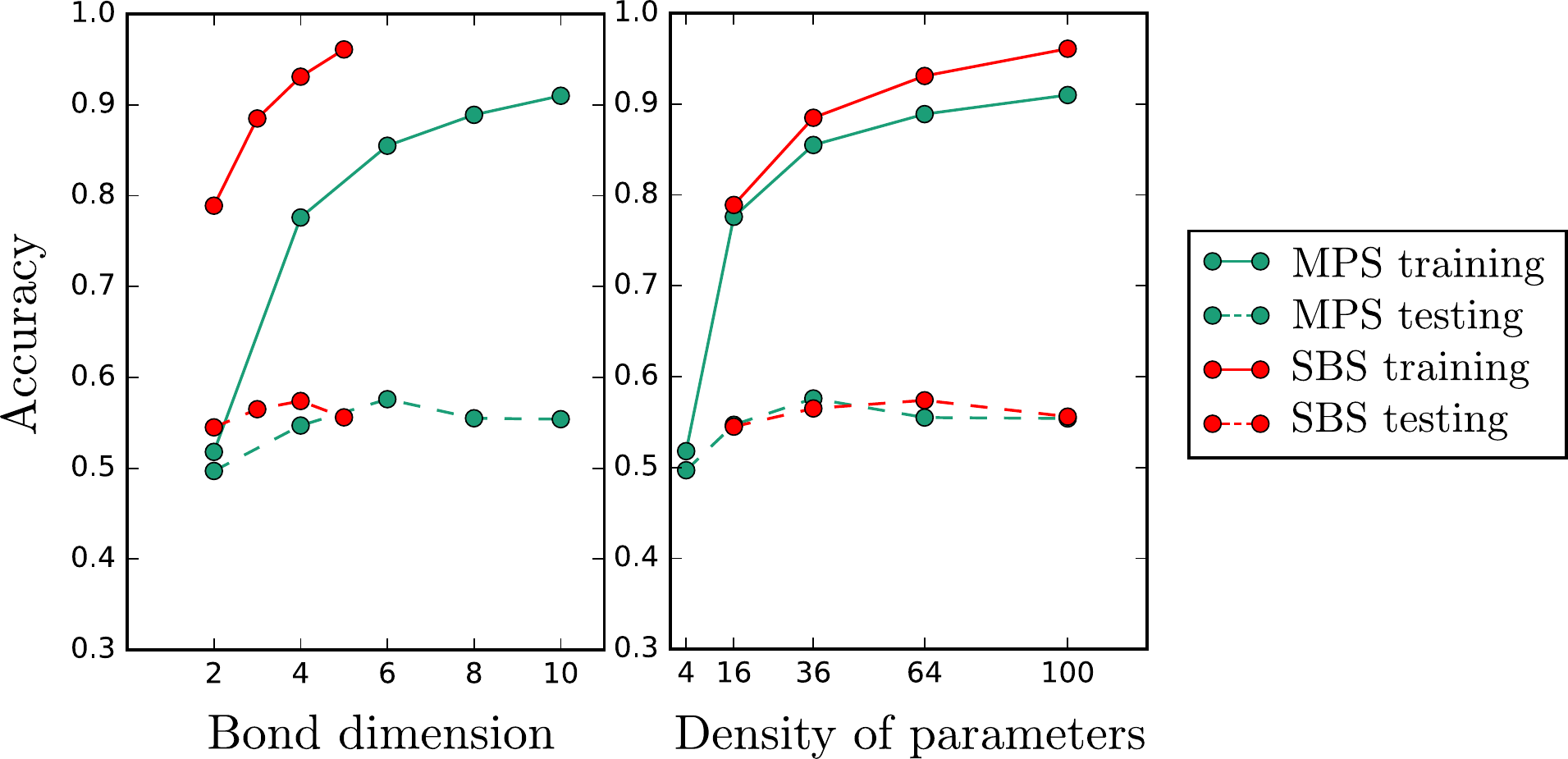}
 \caption{Training and testing accuracy of a MPS and a SBS with 4 strings on the UrbanSound8K data set. The density of parameters is the total number of parameters divided by $174$ (the length of the strings).}
 \label{soundaccuracy}
 \end{figure}

So far we have considered black and white images, but it is also interesting to study how generalized tensor networks could be used for other types of data. In the following we consider the task of classifying environmental sounds. The UrbanSound8K data set \cite{Salamon2014} is a collection of 8732 audio clips (4s or less) divided into 10 classes of urban sounds: air conditioner, car horn, children playing, dog barking, drilling, engine idling, gun shot, jackhammer, siren and street music. The data set is divided into 10 folds and we use folds 1-9 for training and fold 10 for testing. The one-dimensional structure of sounds allows us to compare MPS and SBS with the same 1D string geometry. Preprocessing of the data takes place as follows : clips shorter than 4s are repeated to reach a fixed length of 4s. The first 13 Mel-frequency cepstral coefficients (MFCCs) are extracted for each clip (sampled at 22050Hz) using a window size of 2048 and hop length of 512, resulting in a sequence of length 173 and dimension 13 (Fig.~\ref{sound}). The corresponding 13-dimensional vectors are used as input feature vectors for the tensor network, and the time dimension of the sequence corresponds to the 1-dimensional structure of the MPS, or the strings of the SBS. Note that we do not perform any data augmentation nor split the training examples to enlarge the size of the data set, since we are interested in comparing MPS and SBS, rather than achieving the best possible accuracy on this data set. The training and testing accuracies are reported in Fig.~\ref{soundaccuracy} for a MPS with bond dimension up to 10 and a SBS with 4 strings and bond dimension up to 5. Since we are interested in comparing the expressivity of the different networks, no regularization is used and training is performed until the training accuracy does not improve anymore. Note that a MPS with bond dimension D has as many variational parameters as a SBS with 4 strings and bond dimension $D/2$.

We observe that the SBS has slightly higher training accuracy than a MPS with larger bond dimension and the same number of parameters. The test set performance is not significantly different between distinct architectures and in both cases we find that a lot of overfitting has taken place, which is not surprising given the small number of training examples. Higher accuracies have been reported with other methods on the same data set. For example Convolutional Neural Networks can reach above $70\%$ test set accuracy \cite{Piczak2015}, however they use many more input features and rely on data augmentation. Nevertheless our results show that SBS should also be considered along with MPS when considering one-dimensional data, and may be applied in other settings such as natural language processing \cite{Pestun2017a,Pestun2017b}.

\bigskip
\section{Conclusion}
\label{SEC:Conclusion}
\bigskip
We have introduced generalized tensor networks, which enlarge the class of tensor networks by introducing a reuse of information taking the form of a copy operation of tensor elements. The resulting networks are related to graphical models and we have discussed the strong relations that exist between particular graphical models and tensor network structures, such as restricted Boltzmann machines and string-bond states. We provided an algorithm to train these models to perform a supervised learning task and discussed several strategies to use tensor networks in conjunction with real-valued data. We showed that generalized tensor networks that can be contracted exactly can perform accurate image classification with much smaller bond dimension than regular tensor networks, that they can be used in other settings such as sound recognition and that they can be combined with neural-network architectures. Tensor networks can also be seen as a tool to simulate quantum circuits, and there is much research trying to understand how quantum circuits can be used in machine learning. Quantum circuits corresponding to MPS or tree tensor networks have been studied in the context of quantum machine learning \cite{Stoudenmire2016,Huggins2018}. To implement the function corresponding to a generalized tensor network, one would need to copy the data at the input of the quantum circuit. Our results thus show that quantum machine learning circuits should take as input several copies of each data input and not just a single one, as well as share parameters between the different unitaries in the circuit. Generalized tensor networks which originate from the classical simulation of quantum states may thus serve as a testing and benchmarking platform of near-term quantum machine learning architectures.

\section*{Acknowledgment}

We would like to thank Shi-ju Ran, Yoav Levine and Miles Stoudenmire for discussions, as well as Vedran Dunjko for a careful reading of the paper and helpful comments. This work was supported by the ERC Advanced Grant QENOCOBA under the EU Horizon 2020 program (grant agreement 742102) and the German Research Foundation (DFG) under Germany's Excellence Strategy through Project No.\ EXC-2111 - 390814868 (MCQST). N.P. acknowledges financial support from ExQM.

\bibliography{bibliorbm}

\end{document}